\documentclass[manuscript]{aastex}

\slugcomment{Not to appear in Nonlearned J., 45.}

\shorttitle{K-Shell Photoabsorption of Magnesium Ions}
\shortauthors{Haso{\u g}lu et al.}

\begin{document}

\title{K-Shell Photoabsorption of Magnesium Ions}

\author{M. F. Haso{\u g}lu}
\affil{Department of Computer Engineering, Hasan Kalyoncu University, 27100 Sahinbey, Gaziantep, Turkey}

\author{Sh. A. Abdel-Naby}
\affil{Department of Physics, Auburn University, Auburn, Alabama 36849, USA}

\author{E. Gatuzz}
\affil{Centro de F\'{i}sica, Instituto Venezolano de Investigaciones Cient\'{i}ficas, Caracas 1020, Venezuela}

\author{J. Garc\'{i}a}
\affil{Harvard-Smithsonian Center for Astrophysics, MS-6, 60 Garden Street, Cambridge, MA 02138, USA}

\author{T.~R. Kallman}
\affil{NASA Goddard Space Flight Center, Greenbelt, MD 20771, USA}

\author{C. Mendoza}
\affil{Department of Physics, Western Michigan University, Kalamazoo, MI 49008-5252, and Centro de F\'{i}sica, Instituto Venezolano de Investigaciones Cient\'{i}ficas, Caracas 1020, Venezuela}

\author{T. W. Gorczyca}
\affil{Department of Physics, Western Michigan University, Kalamazoo, MI 49008-5252, USA}

\begin{abstract}
  X-ray photoabsorption cross sections have been computed for all magnesium ions using the $R$-matrix method. A comparison with the other available data for \ion{Mg}{2}--\ion{Mg}{10} shows good qualitative agreement in the resultant resonance shapes. However, for the lower ionization stages, and for singly-ionized \ion{Mg}{2} in particular, the previous $R$-matrix results \citep{witt,witt2} overestimate the K-edge position due to the neglect of important orbital relaxation effects, and a global shift downward in photon energy of those cross sections is therefore warranted. We have found that the  cross sections for \ion{Mg}{1} and \ion{Mg}{2} are further complicated by the M-shell ($n=3$) occupancy. As a result, the treatment of spectator Auger decay of $1s\rightarrow np$ resonances using a method based on multichannel quantum defect theory and an optical potential becomes problematic, making it necessary to implement an alternative, approximate treatment of Auger decay for neutral \ion{Mg}{1}. The new cross sections are used to fit the Mg K edge in {\it XMM-Newton} spectra of the low-mass X-ray binary GS 1826-238, where most of the interstellar Mg is found to be in ionized form.
\end{abstract}


\section{Introduction}

Accurate K-shell photoabsorption cross sections are necessary for modeling astrophysical plasmas, interpreting the observed spectra from distant cosmic emitters, and determining the elemental abundances of the interstellar medium (ISM). Spectra of K-shell processes can be observed from all ionic stages of the most abundant elements between oxygen and nickel \citep{Paerels03}. Magnesium K lines, in particular, have been observed since the early days of X-ray astronomy; for instance, in the spectrum of the O4 $\zeta$~Puppis star taken with the Broad-Band X-Ray Telescope  associated with a thermal plasma of $\sim 6\times 10^6$~K \citep{cor93}. These findings were used, in spite of observational shortcomings such as short exposure times, to constrain the location, temperature, and chemical abundances of the emitting region.

With the advent of high-resolution satellite-borne observatories such as {\em Chandra}, {\em XMM-Newton}, and {\em Suzaku}, X-ray spectroscopy has come of age to provide unique opportunities for studying in detail the physical conditions and processes of exotic and energetic sources. Such is the case of emission K~lines from hydrogen-, helium-, and lithium-like Mg ions observed in the supernova remnant N132D \citep{beh01}; the active nucleus of the giant elliptical galaxy M~87 \citep{sak02}; the massive X-ray binary Cen~X-3, where the importance of resonant line scattering in photoionized plasmas was highlighted \citep{woj03}; the accretion disk of the low-mass X-ray binary system EXO 0748-676  providing evidence of a neutron star \citep{jim03}; the Seyfert~2 galaxy Mkn~3 \citep{pou05}; the outflow component of quasar PG 1211+143 \citep{kas06}; the Cap region above the disk of the starburst galaxy M82, where metal abundance ratios are consistent with Type-II supernova nucleosynthesis \citep{tsu07}; the Fornax intracluster medium allowing an accurate measurement of the Mg abundance that leads to an estimate of the ratio $[{\rm Mg/Fe}]\approx 0.5{-}0.7$ that reflects a stellar metalicity \citep{mat07}; the highly variable narrow-line Seyfert Type 1 galaxy NGC 4051 giving outflow velocities of $\sim 600$~km\,s$^{-1}$ \citep{lob11}; and the Galactic supernova remnant G346.6-0.2 that suggests ejecta-dominated emission with relative abundances pointing to a Type Ia supernova explosion \citep{sez11}. \citet{yam12} have recently studied the spectrum of the Galactic supernova remnant G344.7–0.1, detecting for the first time in an extended celestial source the K$\alpha$ line from \ion{Al}{12} at $\sim 1.6$~keV. This is an important finding because, since both Mg and Al are synthesized during C/Ne burning, the Al/Mg abundance ratio would be a sensitive metalicity diagnostic. This indicator is currently limited by both spectral resolution, which will be improved in the near future with the launching of the {\em Astro-H} telescope, and atomic data.

Absorption Mg K lines are also observed in ISM spectra towards X-ray sources where a desirable feature would be to determine the amount of this element locked up in grain minerals. \citet{pinto} have measured the spectrum of the low-mass X-ray binary GS 1826-238, finding a Mg abundance of $2.45\pm 0.35$ solar that appears to be consistent with a proposed Galactic gradient. Also, the ultra-compact binary candidate 4U0614+091 observed by \citet{schulz} shows a strong variability that causes an excess component intrinsic to the source that demands adjustments of the Ne edge; however, an excess in the Mg edge is not as yet conclusive partly due to its unestablished morphology.

In previous studies, we have carried out accurate calculations of K-shell photoabsorption cross sections that have been applied to X-ray spectral diagnostics; e.g. all ionization stages of carbon \citep{Hasoglu10}, oxygen \citep{goro,garo,juetto}, and neon \citep{gorne,juettne}. This project is hereby extended to the magnesium isonuclear sequence where a further complication arises for the two lowest ionization stages (\ion{Mg}{1}  and \ion{Mg}{2}) as the $n=3$ M-shell becomes occupied: the atomic radius now doubles in size, and our usual treatment based on multichannel quantum defect theory (MQDT) and an optical potential \citep{Gorczyca99} becomes problematic for the lower $1s\rightarrow np$ resonances. We have nevertheless come up with an approximate procedure for treating these cases that yields reliable X-ray photoabsorption cross sections for all relevant magnesium ions, i.e. \ion{Mg}{1}--\ion{Mg}{10}. Furthermore, these new cross sections will allow us to revise the Mg abundance in the low-mass X-ray binary GS 1826-238 and, in particular, to determine the ionic fractions, if any, of the lowly ionized species.

\section{Theoretical Approach}

K-shell photoabsorption consists of the direct photoionization of the $1s$ electron, which is treated in a straightforward manner using $R$-matrix methods, and the strong $1s \rightarrow np$ photoabsorption resonances. Photoexcitation of these resonance states is then followed by two competing decay routes. The first is {\em participator} Auger decay, where the $np$ valence electron takes part in the autoionization process with a decay rate that scales as $1/n^3$ and goes to zero near the K-shell threshold. These channels are included in the $R$-matrix calculation. The second route is {\em spectator} Auger decay, in which $np$ the valence electron is oblivious to the autoionization process giving instead a decay width that is independent of $n$. Therefore, the latter is the dominant decay route as $n\rightarrow \infty$, and guarantees a smooth cross section as the K-shell threshold is approached. Above threshold, K-shell photoionization to the $1s2\ell^q$ states occurs instead.

For the present work we use the $R$-matrix method \citep{berrington95,burke} with modifications to account for the  spectator Auger broadening via an optical potential as described by \citet{Gorczyca99}. This enhanced $R$-matrix method has been shown to be accurate in describing experimental synchrotron measurements for argon  \citep{Gorczyca99}, oxygen \citep{goro}, neon \citep{gorne}, and carbon \citep{Hasoglu10}. The Auger widths for the $1s2\ell^q$ states are computed by applying the \citet{smith}  time-delay method to the photoabsorption $R$-matrix calculation of the neighboring $1s^22\ell^{q-1}$ magnesium ion. Further details can be found in our previous work \citep{Gorczyca99, goro, gorne, Hasoglu10}.

\section{Cross Section Results}
\label{sec:results}

As an assessment of the present atomic description, the computed target-state energies and binding energies are presented in Tables~\ref{mgIen}--\ref{mgXen}, which show fairly good agreement with the recommended NIST spectroscopic values. The computed core Auger widths, which are used within the MQDT optical potential approach for treating spectator Auger resonance broadening, are listed in Tables~\ref{mgIaw}--\ref{mgIXaw}. Comparison with other available data shows fairly good agreement in most cases, indicating again that the present atomic representations are sufficient. It is worth mentioning that the spectator Auger width used in our calculations only changes the shape of the resonance absorption profile, not the strength.

The present K-shell photoabsorption cross sections are shown in Figs.~\ref{mgIpa}--\ref{mgXpa} where the independent-particle (IP) photoionization results of \citet{Verner93} are also included. It can be seen that the present $R$-matrix results are in good quantitative agreement with the IP results above the K-shell thresholds, but the IP cross sections lack the important resonance absorption lines below threshold.

\subsection{Neutral Mg}

\begin{figure}[hbtp]
\centering
\includegraphics[width=4.5in,angle=-90]{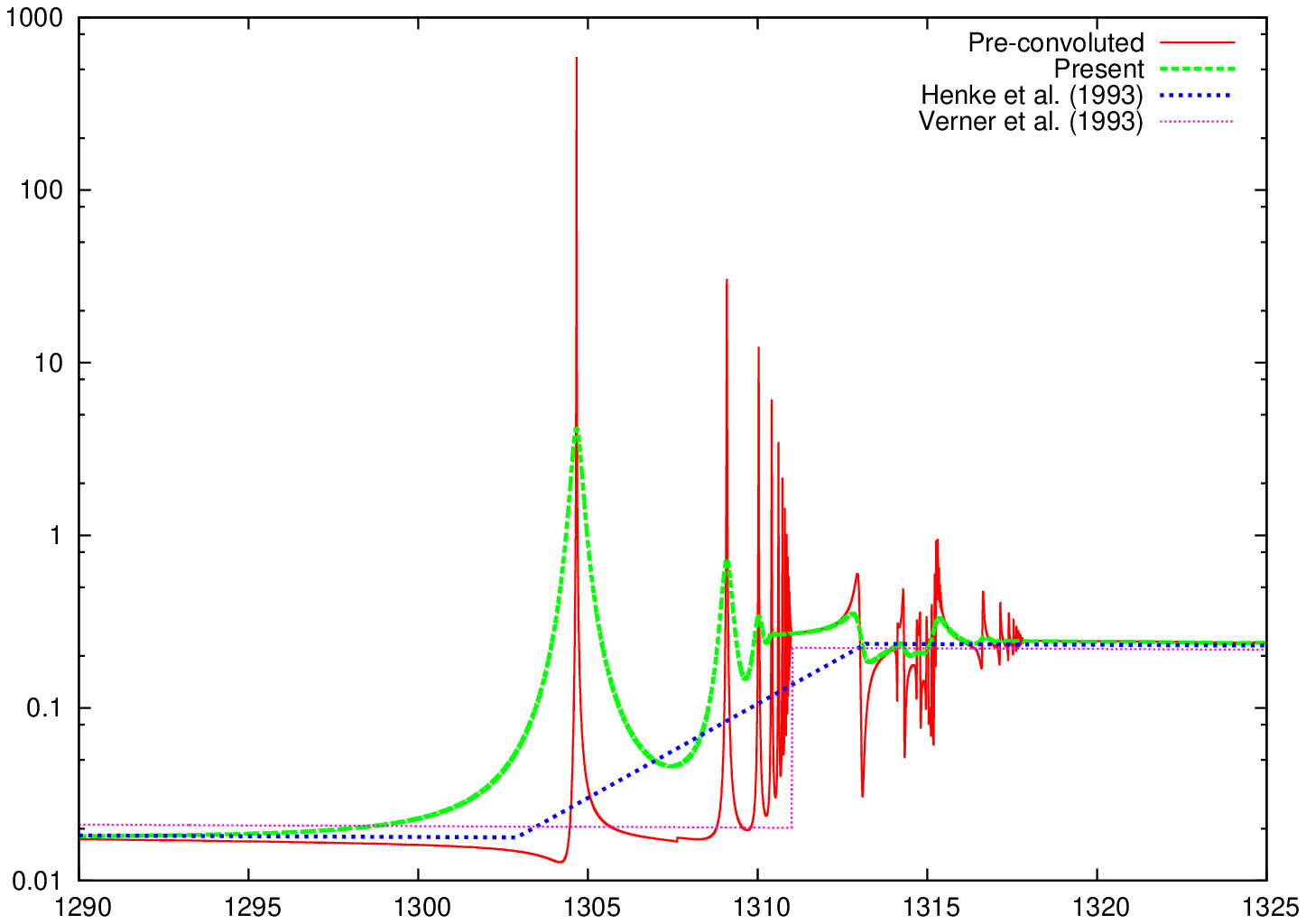}
\caption{\ion{Mg}{1} photoabsorption cross sections. Pre-convoluted cross section obtained by damping the core excited states (see Table~\ref{mgIen}) with an artificially smaller 0.001 Ryd spectator Auger width, then performing a Lorentzian convolution with the calculated Auger-width of $0.0254$~Ryd for the $1s^22s^22p^63s^2\,^2S$ state (see Table~\ref{mgIaw}) to get the final cross section.   Also shown are the IP results \citep{Verner93} and solid-state experimental results \citep{Henke93}.}  \label{mgIpa}
\end{figure}

\begin{figure}[!h]
  \centering
  \includegraphics[width=4.5in,angle=-90]{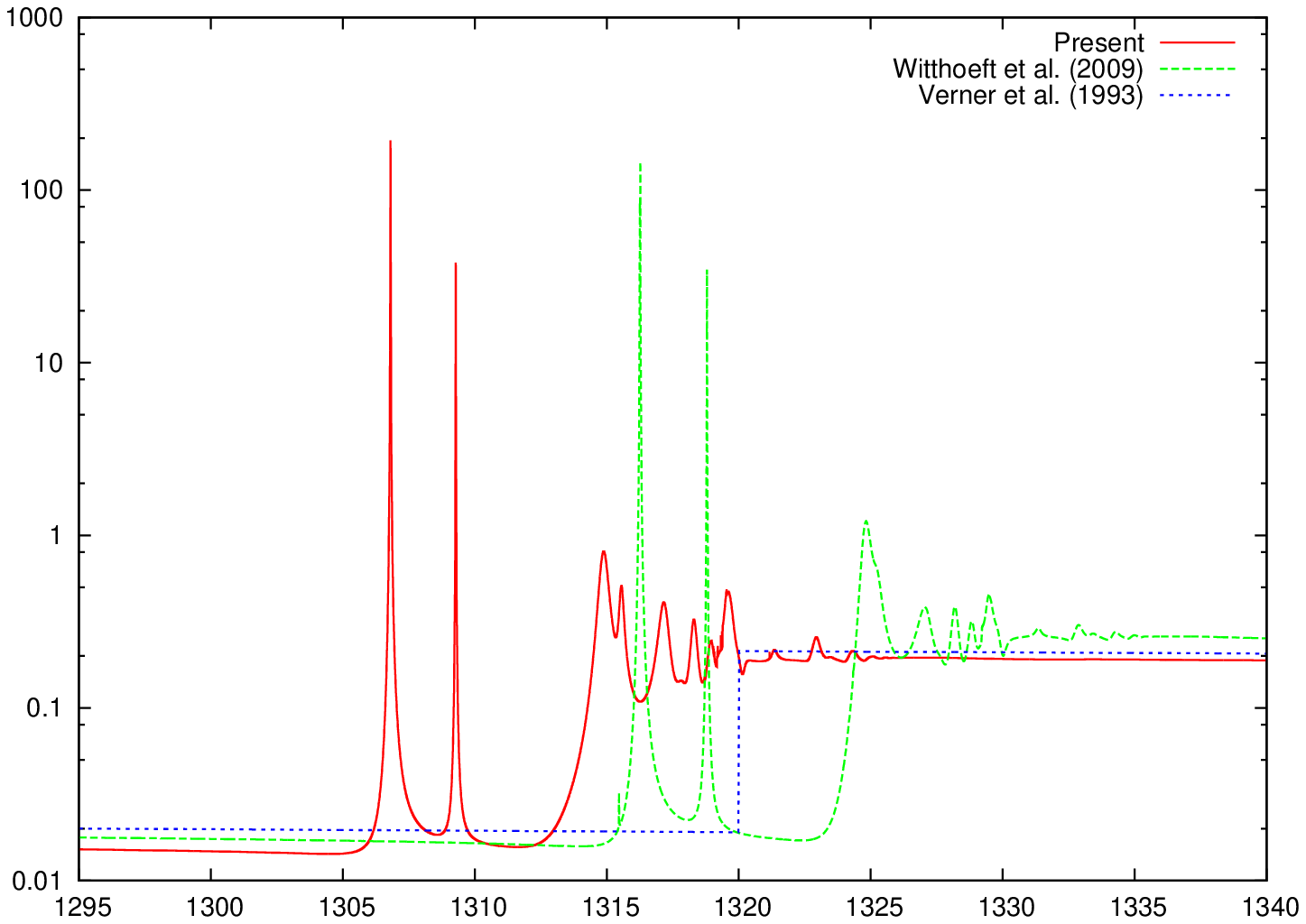}
  \caption{\ion{Mg}{2} photoabsorption cross sections: present results compared to earlier $R$-matrix results \citep{witt2} and IP results \citep{Verner93}.} \label{mgIIpa}
\end{figure}

\begin{figure}[!h]
  \centering
  \includegraphics[width=4.5in,angle=-90]{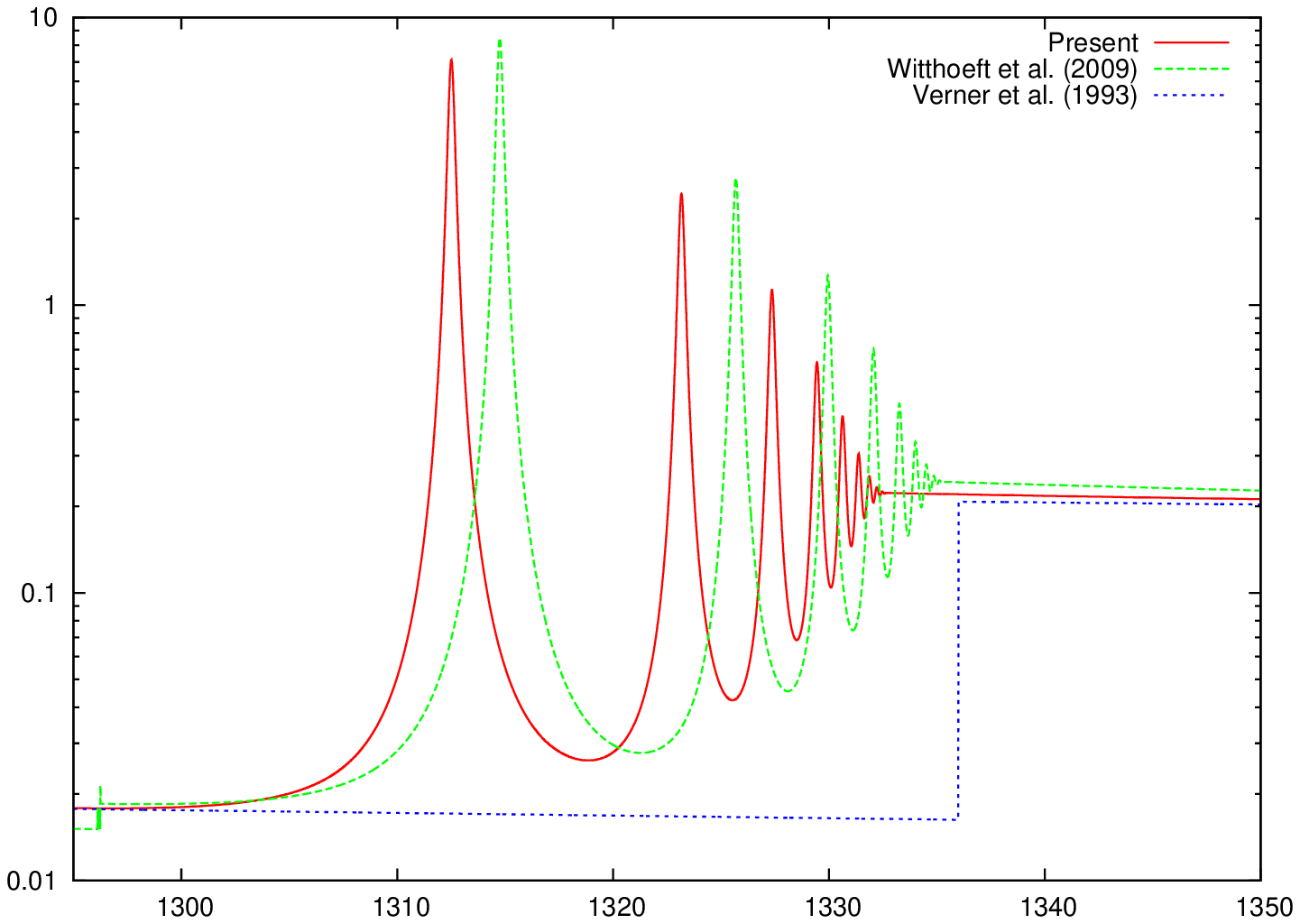}
  \caption{\ion{Mg}{3} photoabsorption cross sections: present results compared to earlier $R$-matrix results \citep{witt} and IP results \citep{Verner93}.} \label{mgIIIpa}
\end{figure}

\begin{figure}[!h]
  \centering
  \includegraphics[width=4.5in,angle=-90]{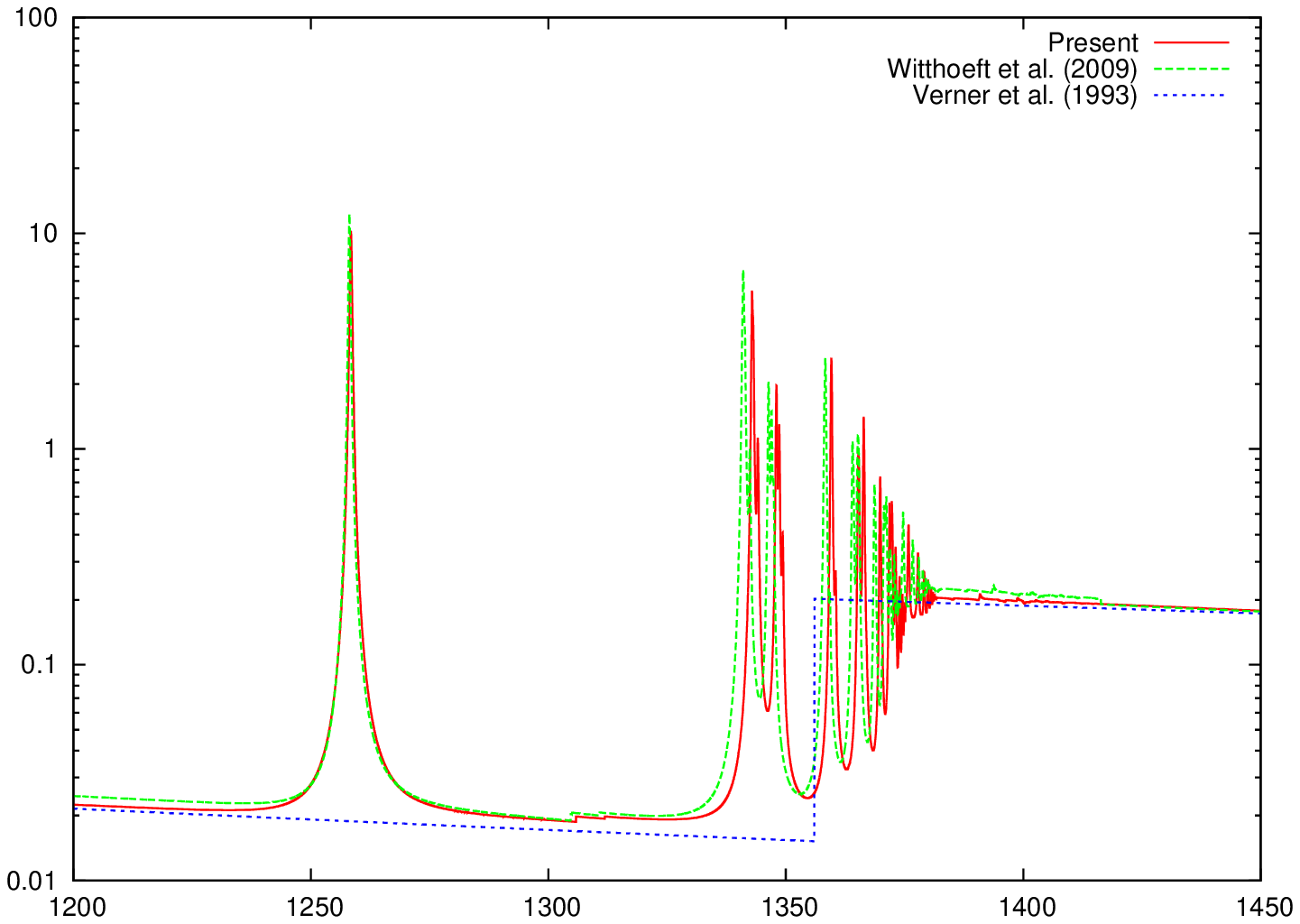}
  \caption{\ion{Mg}{4} photoabsorption cross sections: present results compared to earlier $R$-matrix results \citep{witt} and IP results \citep{Verner93}.} \label{mgIVpa}
\end{figure}

\begin{figure}[!h]
  \centering
  \includegraphics[width=4.5in,angle=-90]{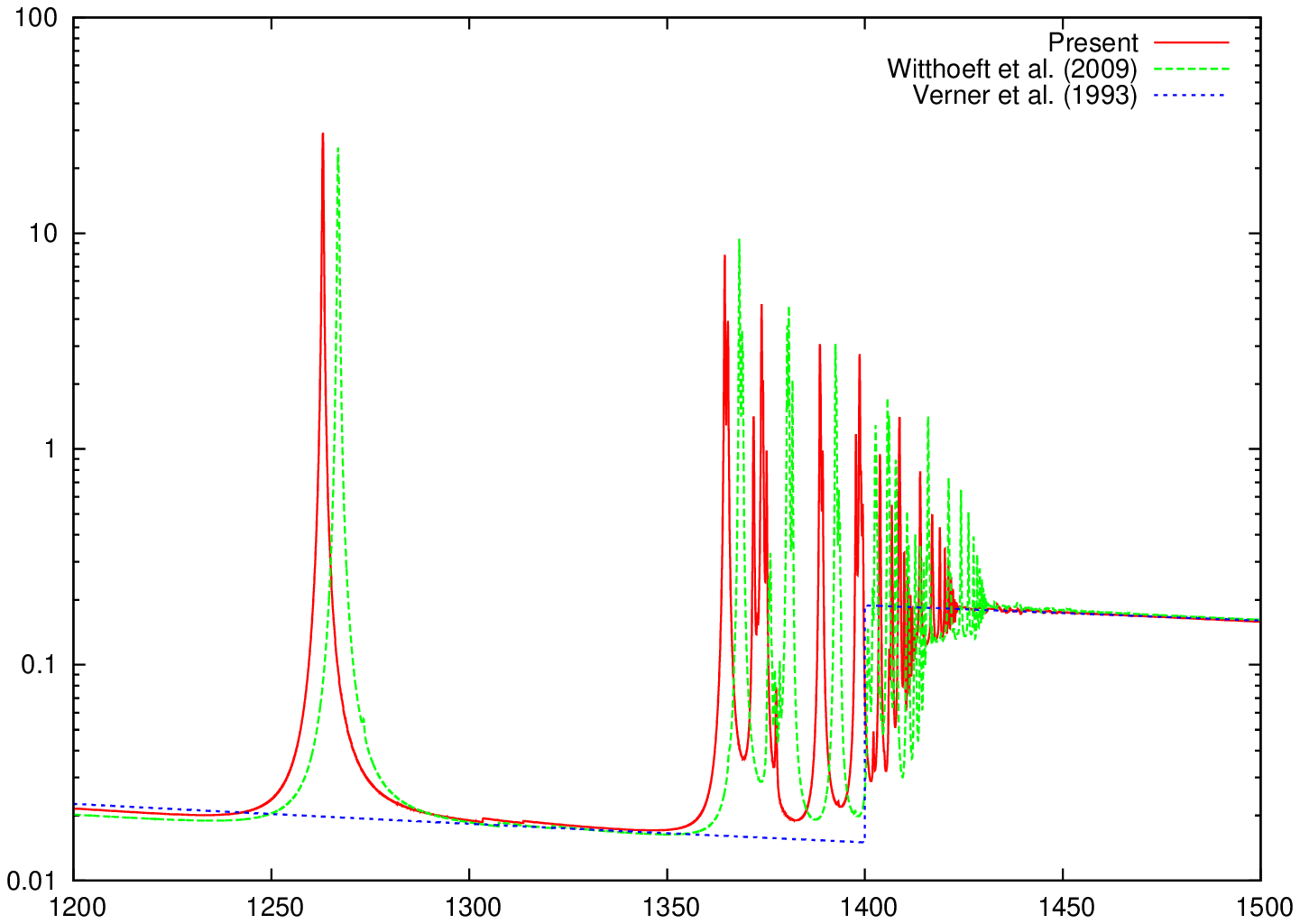}
  \caption{\ion{Mg}{5} photoabsorption cross sections: present results compared to earlier $R$-matrix results \citep{witt} and IP results \citep{Verner93}.} \label{mgVpa}
\end{figure}

\begin{figure}[!h]
  \centering
  \includegraphics[width=4.5in,angle=-90]{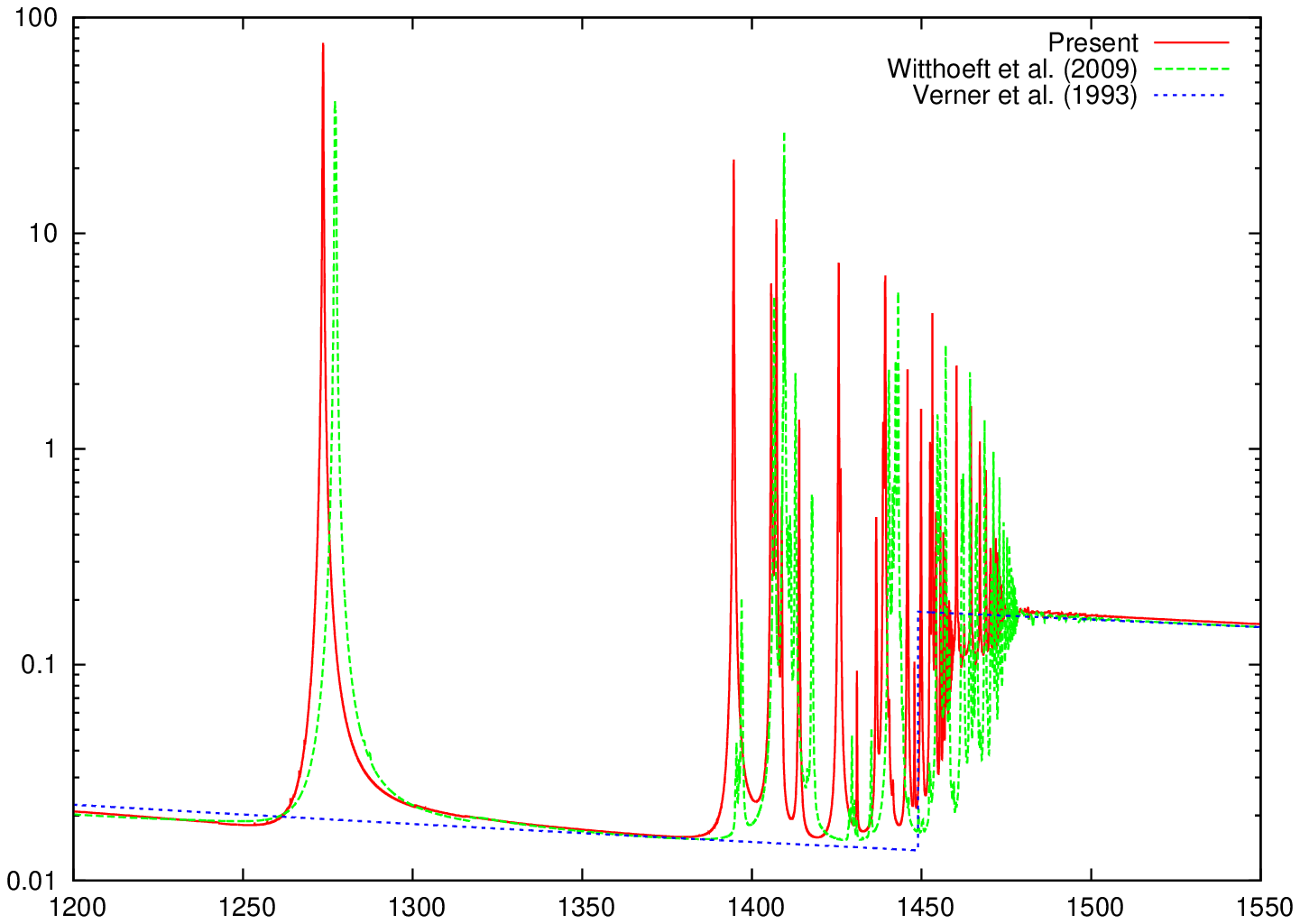}
  \caption{\ion{Mg}{6} photoabsorption cross sections: present results compared to earlier $R$-matrix results \citep{witt} and IP results \citep{Verner93}.} \label{mgVIpa}
\end{figure}

\begin{figure}[!h]
  \centering
  \includegraphics[width=4.5in,angle=-90]{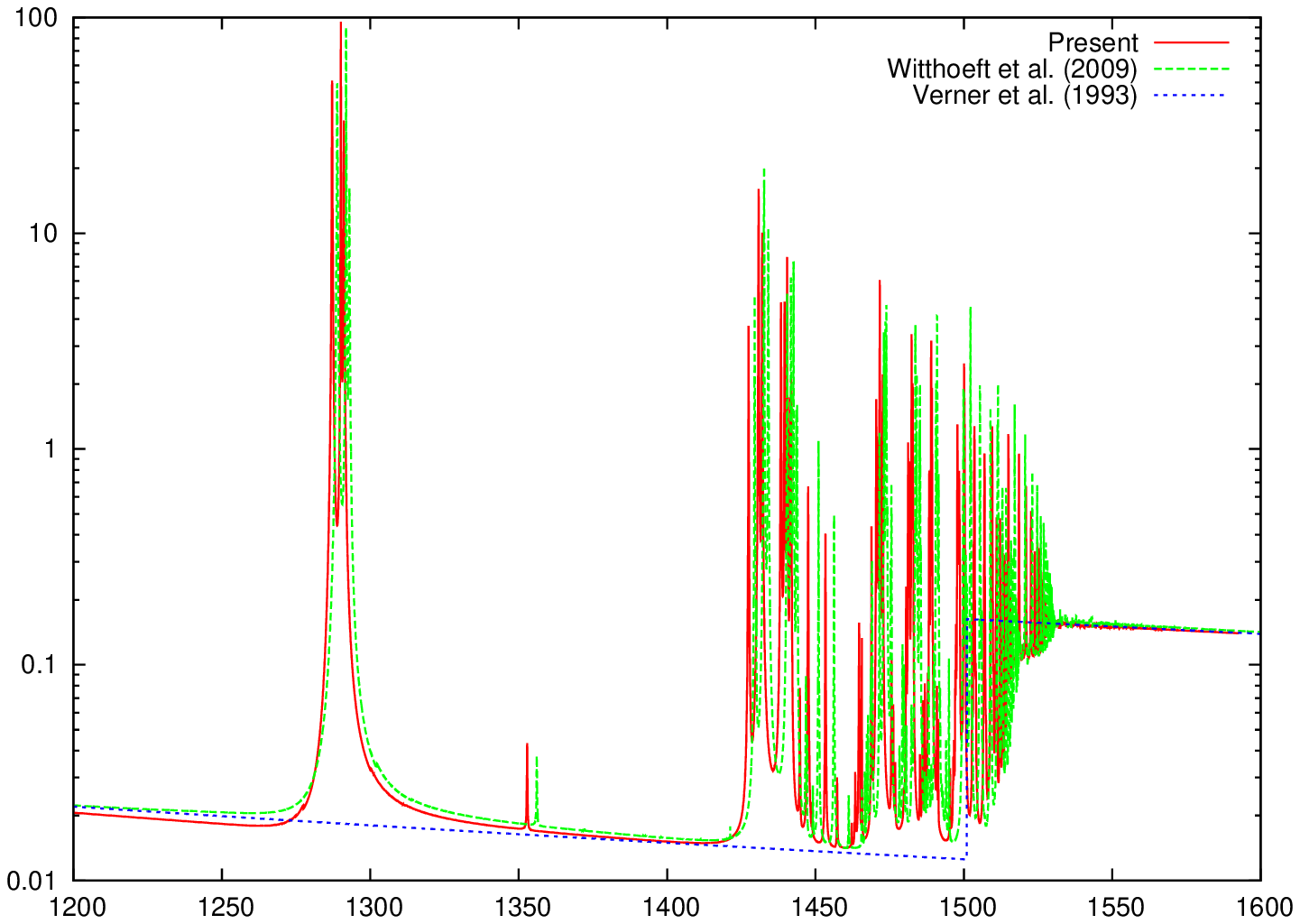}
  \caption{\ion{Mg}{7} photoabsorption cross sections: present results compared to earlier $R$-matrix results \citep{witt} and IP results \citep{Verner93}.} \label{mgVIIpa}
\end{figure}

\begin{figure}[!h]
  \centering
  \includegraphics[width=4.5in,angle=-90]{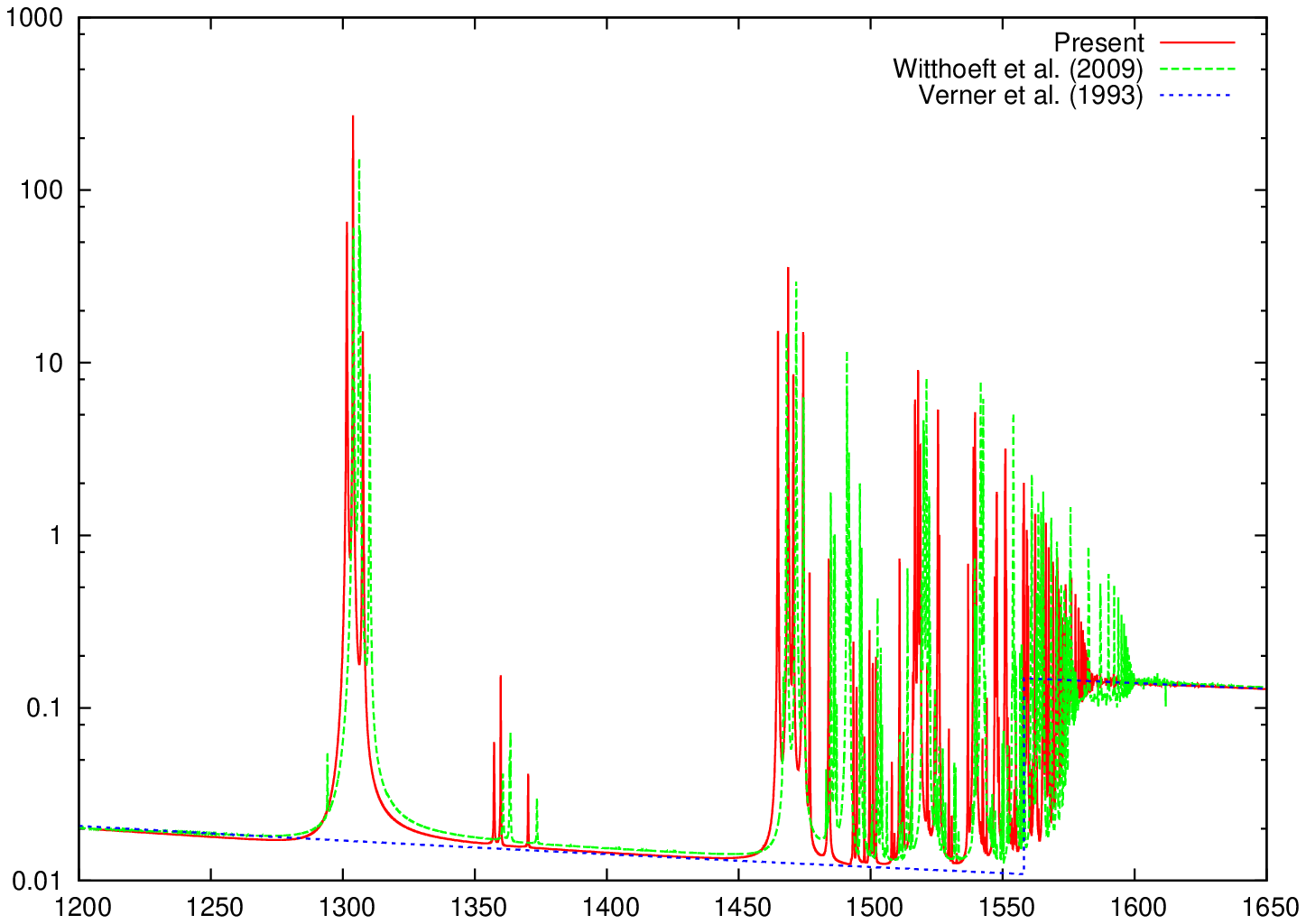}
  \caption{\ion{Mg}{8} photoabsorption cross sections: present results compared to earlier $R$-matrix results \citep{witt} and IP results \citep{Verner93}.} \label{mgVIIIpa}
\end{figure}

\begin{figure}[!h]
  \centering
  \includegraphics[width=4.5in,angle=-90]{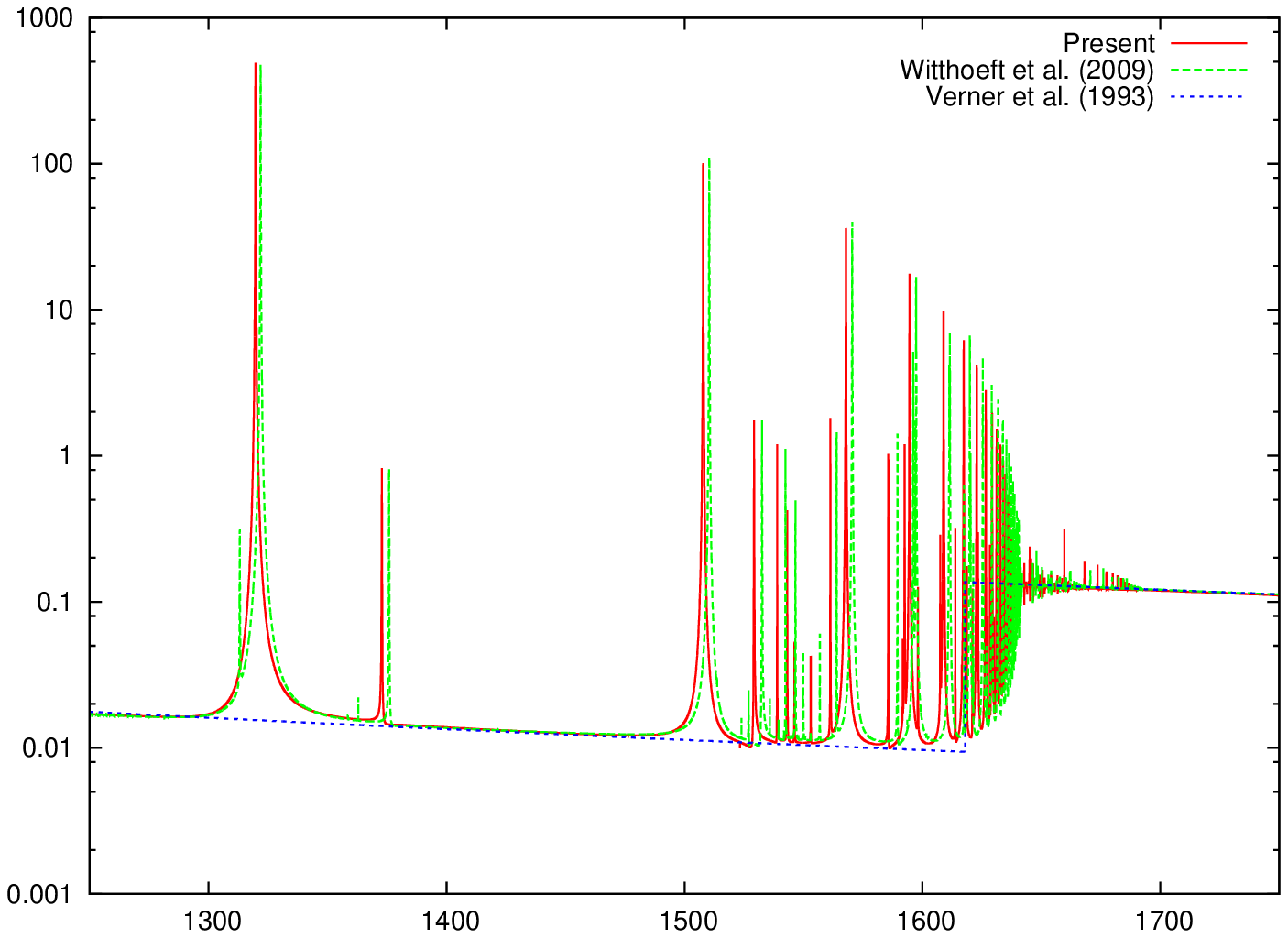}
  \caption{\ion{Mg}{9} photoabsorption cross sections: present results compared to earlier $R$-matrix results \citep{witt} and IP results \citep{Verner93}.} \label{mgIXpa}
\end{figure}

\begin{figure}[!h]
  \centering
  \includegraphics[width=4.5in,angle=-90]{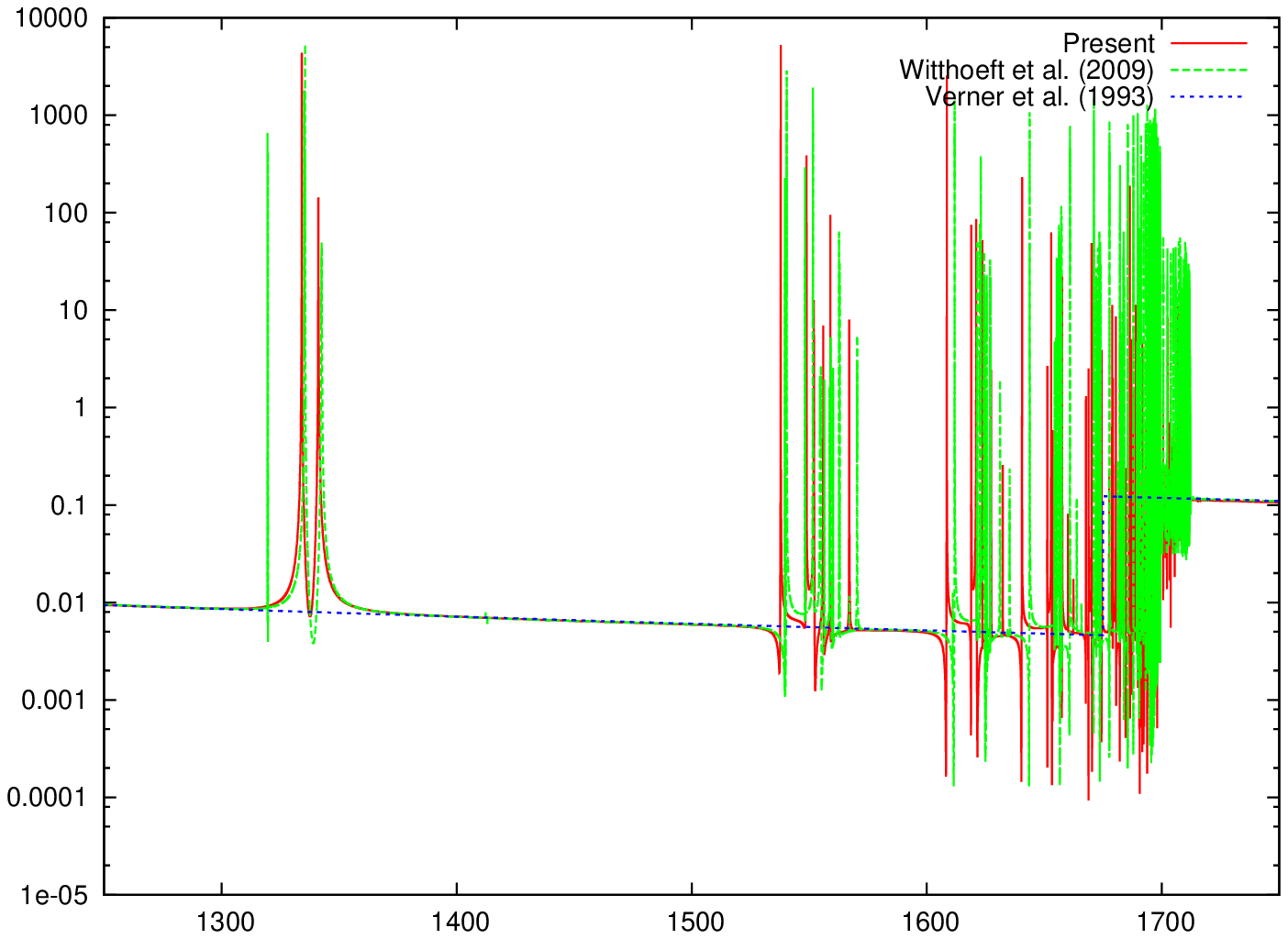}
  \caption{\ion{Mg}{10} photoabsorption cross sections.  Present results compared to earlier $R$-matrix results \citep{witt} and IP results \citep{Verner93}.} \label{mgXpa}
\end{figure}

For the \ion{Mg}{1} photoabsorption cross section (Fig.~\ref{mgIpa}), there are no other $1s\rightarrow np$ resonance cross sections available for comparison. However, as can be deduced from Table~1, our theoretical K-shell threshold is at 1311.03~eV
in fairly good agreement with the experimental value of 1311.4~eV \citep{mgexp}. Furthermore, since our above-threshold cross section is seen to align with the IP results, we are confident about the below-threshold resonance oscillator strength that merges to the above-threshold oscillator strength density through continuity conditions intrinsic within the $R$-matrix framework. Also shown in Fig.~\ref{mgIpa} are the solid-state experimental results of \citet{Henke93}; it is interesting to note that the present $R$-matrix results for {\em neutral} magnesium align more closely with experiment than with the IP results which do not include relaxation effects but shift the threshold downward to align with experiment.

Regarding the $R$-matrix calculations for \ion{Mg}{1}, it was not possible to apply the usual MQDT optical potential method \citep{Gorczyca99} to render spectator Auger broadening since, due to the larger radius of the $R$-matrix box  \citep{berrington95, burke}, the energy dependence of the MQDT parameters at the lower resonances invalidated the simple $E\rightarrow E+i\Gamma/2$ substitution \citep{Gorczyca99}. Instead, a more rigorous approach (beyond the scope of this paper)
is necessary for the adequate modeling of the Auger width. (As noted before, the strength is not affected by the particular width used.) In order to present reliable cross sections for this study, then, we first
perform calculations using a spectator width that is small enough such that the energy-dependent MQDT parameters can still be treated as constant over the width of a resonance, but large enough such that the resonance Rydberg series can be mapped out with a finite number of $R$-matrix energy points. These cross sections are then further convoluted with a Lorentzian profile of width 0.0254~Ryd (see Table 11) to simulate the known Auger broadening.

\subsection{Ionized Mg species}

The only other reported K-absorption ($1s\rightarrow np$) cross sections of Mg ionized species, to our knowledge, are those by \citet{witt} and \citet{witt2}, which have been computed with a similar $R$-matrix approach; however, important orbital relaxation effects were therein neglected. Relaxation is due to the sudden change in the potential perceived by the outermost electrons following excitation or ionization of an inner-shell electron ($1s$ electron in this case), the relative change in potential strength reaching a maximum at the lower ionization stages; hence, \ion{Mg}{2} is expected to be the most affected by relaxation as evidenced by the K-shell threshold being overestimated by approximately 10~eV. This overestimate, due to the absence of relaxation effects, is seen to diminish as the ionic charge increases to the order of 2~eV, which seems to indicate lack of correlation perhaps from strong $2p^2\rightarrow \overline{3d}^2$ double promotions that, in addition, would require the inclusion of optimized $\overline{3d}$ pseudo-orbitals.

\subsection{Final Atomic Data}

Having computed reliable photoabsorption cross sections for \ion{Mg}{1}-\ion{Mg}{10} in the vicinity of their respective K-edges, we then produced final data sets, to be used in the {\sc xstar} spectral modeling code \citep{bau01}, by a single fitting formula for each ion, as described more fully in a recent paper on \ion{O}{1} \citep{gor13}.  Briefly, the X-ray photoabsorption cross section for each ion is modeled as a sum of contributions from all 
possible photoionization mechanisms.
For the direct (non-resonant) photoionization cross sections of the $1s$, $2s$, $2p$, and (for \ion{Mg}{1} and \ion{Mg}{2}) the $3s$ sub-shells, the analytic 
formulas given by \citet{Verner93,ver96} are used.  However, due to the relatively larger relaxation effects for neutral \ion{Mg}{1}, we use instead a three-parameter asymptotic
inverse power law fit to the data of \citet{Henke93} for the $1s$ partial cross section; a similar procedure was done for neutral \ion{O}{1}, for reasons discussed at length in that paper \citep{gor13}.  

While the \citet{Verner93,ver96} or inverse power formulas yield the direct partial cross sections, the resonant absorption cross sections -
predominantly due to the $1s\rightarrow np$ resonances - that contribute to the total 
cross section are represented by infinite sums of Lorentzian profiles for each Rydberg series.  We note also that for the weaker, excited $1s2s^22p^63s3pns$ and $1s2s^22p^63s3pnd$ series in  \ion{Mg}{1}, as shown above the \ion{Mg}{2} $1s2s^22p^63s^2$ threshold of 1311 eV in Fig.~1, a modified, asymmetric Fano for the profile is used rather than a Lorentzian fit.  This single analytic expression for the total photoabsorption cross section - one for each Mg ion, differing only in the fitting parameters used -
ensures a reliable and continuous data set for the entire x-ray region of interest, going well above and well below the Mg K-edge region.

\section{Modeling the Mg K edge}\label{sec_fit}

We use {\it XMM-Newton} spectra from the low-mass X-ray binary GS 1826-238 (Galactic coordinates $l=9.27$ and $b=-6.08$) to analyze Mg photoabsorption in the ISM. The data were obtained with the Reflection Grating Spectrometers (RGS) and the reduction process was performed with the {\it XMM-Newton} Science Analysis System (SAS, Version 12.0.1). The two observations (see Table~\ref{obs}) are fitted simultaneously in the 8--11~\AA\ wavelength region, the data being rebinned to obtain at least 20 counts per channel in order to use chi-square statistics \citep{nou89}. For the analysis we have used the {\sc isis} (Version 1.6.2-27) package to compare the {\tt TBnew} and {\tt warmabs} models to estimate the impact of the new atomic data on ISM indices such as ionization state, relative ionic fractions, and elemental abundances. {\tt TBnew} is an X-ray absorption model that includes chemical species from H to Ni by implementing the cross sections of \citet{ver96} although it only considers photoabsorption in neutrals. On the other hand, {\tt warmabs} takes into account both the neutral and all ionized species with $Z\leq30$, and their relative fractions are determined self-consistently by solving the ionization balance in the gas; moreover, its atomic database is being systematically improved and now includes the Mg photoionization cross sections presented here.

\begin{figure}
  \epsscale{1.}
  \plotone{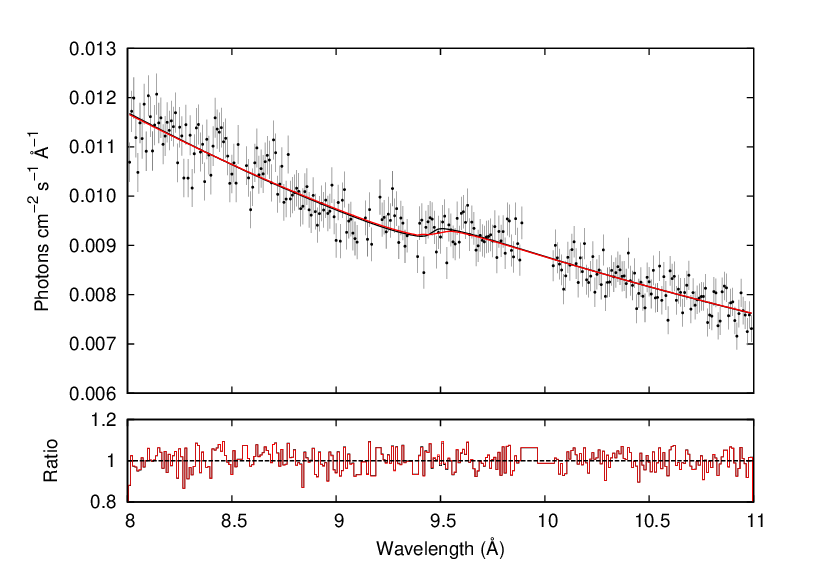}
  \caption{Unfolded  {\it XMM-Newton} RGS spectra of the X-ray binary GS 1826-238 simultaneously fitted in the 8--11~\AA\ region using the {\tt TBnew} (solid black line) and {\tt Warmabs} (solid red line) models. Although the fit is carried out simultaneously, the data was combined for visual purposes.\label{mg_edge}}
\end{figure}

\begin{figure}
  \epsscale{1.}
  \plotone{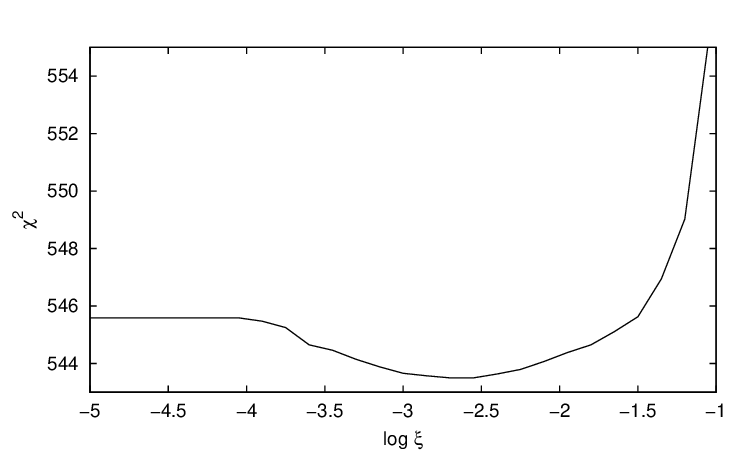}
  \caption{Confidence range for the log $\xi$ parameter. The best-fit corresponds to a low ionization degree but including ionized states (mostly  \ion{Mg}{2} and \ion{Mg}{3}).\label{conf_range}}
\end{figure}

\begin{figure}
  \epsscale{1.}
  \plotone{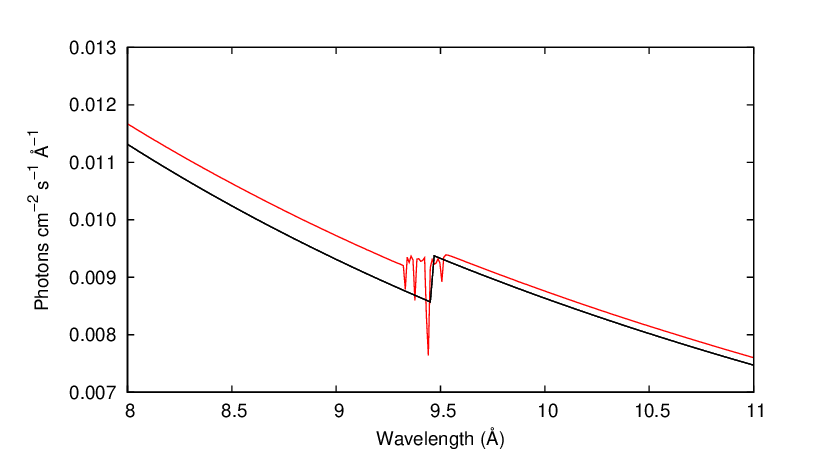}
  \caption{{\tt TBnew} (solid black line) and {\tt Warmabs} (solid red line) model comparison. {\tt TBnew} model only includes neutral magnesium atomic data while {\tt Warmabs} model includes the neutral and all ionized
species.\label{model_comp}}
\end{figure}

Figure~\ref{mg_edge} shows the best fit of both the {\tt TBnew} (solid black line) and {\tt warmabs} (solid red line) models to the {\it XMM-Newton} observations. Although the fit has been carried out simultaneously, the data are combined for visual purposes where the black dots correspond to the observations. The Mg K edge is located at $\approx 9.46$~\AA. The model-to-data ratio shown in the lower panel indicates that both models fit the data satisfactorily, the best-fit parameters for both models being listed in Table~\ref{edge_fit} where abundances are given relative to \citet{lod03}. It may be appreciated that the fit quality ($\chi^2$) for both models is approximately the same although the physical predictions are different: the {\tt TBnew} abundance ($5.24^{+1.68}_{-1.52}$) is 16\% greater than that of {\tt warmabs} ($4.51^{+2.11}_{-1.02}$), and the ionization parameter of the latter ($-2.38^{+0.81}_{-0.97} $) indicates a strong presence of not only \ion{Mg}{1} but also \ion{Mg}{2} and \ion{Mg}{3} ions. A calculation of the confidence region for the ionization parameter is included in Figure~\ref{conf_range} where it may be noted that the best fit requires the presence of ionized states. Figure~\ref{model_comp} shows a comparison of the models in units of flux (photons~cm$^{-2}$~s$^{-1}$~\AA$^{-1}$) in the Mg K-edge region: {\tt TBnew} shows a simple neutral edge while the inclusion of the higher ionized species in {\tt warmabs} results in a more complex and smeared edge. We have derived the Mg column densities using the ionization fractions from the {\tt warmabs} fit (see Table~\ref{col_den}), namely $(6.433 \pm  1.28)\times 10^{16}$~cm$^{-2}$, $(7.085\pm  1.41)\times 10^{16}$~cm$^{-2}$, and $(1.29\pm 0.25)\times 10^{17}$~cm$^{-2}$ for \ion{Mg}{1}--\ion{Mg}{3}, respectively. These values are much larger than those expected in the ISM \citep[e.g.][]{val13} explaining the high abundance in both models; however, the {\tt warmabs} ionization fractions agree with previous UV observations that indicate that Mg is expected to be found predominantly in ionized rather than neutral form \citep{gna06}.

\section{Summary and Conclusion}

We have performed new R-matrix optical potential calculations for the K-shell photoabsorption cross sections of the \ion{Mg}{1}--\ion{Mg}{10} ions.  Comparison with 
IP cross sections \citep{Verner93} indicates that we have computed accurate direct photoionization cross sections away from the resonant region. Concerning the Rydberg series of $1s\rightarrow np$ resonant photoabsorption below threshold, we compare to the only other existing calculations of \citet{witt} and \citet{witt2} for \ion{Mg}{2}--\ion{Mg}{10} (i.e., excluding the all important 
neutral \ion{Mg}{1}), and conclude the following.  Whereas the  cross sections compare well, at least for the background magnitude and the resonance strengths, 
between the two calculations for 
multiply-charged Mg ions, the two begin to differ more 
as the ionization stage is decreased, owing to the increased relaxation effects for low charged systems.  
The earlier R-matrix calculations \citep{witt,witt2} did not include
relaxation effects, and as a result, their K-edge threshold positions were overestimated - by about 10 eV for \ion{Mg}{2}, by about 5 eV for \ion{Mg}{3}, and diminishing to roughly a constant of 2 eV for higher ionization stages.  Furthermore, 

Of equal importance, those earlier R-matrix calculations only treated ionized Mg ions, and no other resonant K-shell photoabsorption cross sections are available for neutral \ion{Mg}{1}.  As a result, the present calculations represent a significant improvement 
in the Mg ion K-shell photoabsorption database, correctly modeling all features in the  vicinity of the K-edge.

These new cross sections have been included in the atomic database of the {\sc xstar} modeling code \citep{bau01} in order to continue the atomic data benchmarking with astronomical spectra currently being carried out by \citet{gat13a, gat13b} and \citet{gor13}. As an initial test, we have fitted the Mg edge in {\it XMM-Newton} spectra of the low-mass X-ray binary GS 1826-238 with models that include both the older IP cross sections of \citet{Verner93} ({\tt TBnew}) and the present cross sections ({\tt warmabs}). Even though these observations are not good enough to give a reliable verdict, an interesting new finding is that most of the Mg happens to be in ionized form; this certainly justifies the present effort to compute accurate cross sections for the ionized species and perhaps explains the abundance difference (a factor of 2) with respect to the value quoted by \citet{pinto}.

\section{Acknowledgment}

This work was supported in part by a NASA APRA grant NNX11AF32G.

\clearpage




\clearpage

\clearpage

\clearpage

\clearpage

\clearpage

\clearpage

\clearpage

\clearpage

\clearpage

\clearpage

\clearpage

\clearpage
\begin{table*}[!hbtp]
  \caption{\label{mgIen} Energies (Ryd) of the \ion{Mg}{2} target states and \ion{Mg}{1} ground state. Also shown are NIST spectroscopic values and HFR1 results \citep{palmeri08}.}\vspace*{0.07in}
  \resizebox{4in}{!}{
  \begin{tabular}{l c c c}  
  \hline \hline
  State                                  & Present   & NIST &  HFR1  \\
  \hline
  $1s^22s^22p^63s^2        \ ^1S$      & $ -0.5428 $ & $ -0.5620 $ & $          $\\[-0.06in]
  \hline
  $1s^22s^22p^63s          \ ^2S$      & $  0.0000 $ & $  0.0000 $ & $  0.0000  $\\[-0.09in]
  $1s^22s^22p^63p          \ ^2P$      & $  0.3162 $ & $  0.3256 $ & $  0.3300  $\\[-0.09in]
  $1s^22s^22p^53s^2        \ ^2P$      & $  3.8240 $ & $  3.6745 $ & $  3.6235  $\\[-0.09in]
  $1s^22s2p^63s^2          \ ^2S$      & $  6.7244 $ & $         $ & $          $\\[-0.09in]
  $1s^2s22p^63s^2          \ ^2S$      & $ 95.8157 $ & $         $ & $  95.8295 $\\[-0.09in]
  $1s^2s22p^63s(^1S)3p     \ ^2P$      & $ 96.1441 $ & $         $ & $  96.1612 $\\[-0.09in]
  $1s2s^22p^63s(^3S)3p     \ ^2P$      & $ 96.3185 $ & $         $ & $  96.5066 $\\[-0.05in]
  \hline
  \end{tabular}
  }
\end{table*}

\clearpage
\begin{table*}[!hbtp]

  \caption{\label{mgIIen}
  Energies (Ryd) of the \ion{Mg}{3} target states and \ion{Mg}{2} ground state.} \vspace*{0.07in}
  \resizebox{4in}{!}{
  \begin{tabular}{l r r}  
  \hline \hline
  State                 & Present & NIST \\
  \hline
  $1s^22s^22p^63s\ ^2S)$  &   $-1.0928 $   & $-1.1051 $ \\[-0.06in]
  \hline
  $1s^22s^22p^6  \ ^1S$  &   $ 0.0000 $   & $ 0.0000 $ \\[-0.09in]
  $1s^22s^22p^53s\ ^3P$  &   $ 3.8897 $   & $ 3.8847 $ \\[-0.09in]
  $1s^22s^22p^53s\ ^1P$  &   $ 3.9384 $   & $ 3.9324 $ \\[-0.09in]
  $1s^22s^22p^53p\ ^3S$  &   $ 4.2561 $   & $ 4.2591 $ \\[-0.09in]
  $1s^22s^22p^53p\ ^1P$  &   $ 4.3615 $   & $ 4.3593 $ \\[-0.09in]
  $1s^22s^22p^53p\ ^3P$  &   $ 4.3624 $   & $ 4.3658 $ \\[-0.09in]
  $1s^22s^22p^53p\ ^1S$  &   $ 4.5778 $   & $ 4.5200 $ \\[-0.09in]
  $1s^22s^22p^53d\ ^3P$  &   $ 4.8345 $   & $ 4.8360 $ \\[-0.09in]
  $1s^22s^22p^53d\ ^1P$  &   $ 4.8821 $   & $ 4.8858 $ \\[-0.09in]
  $1s^22s2p^63s  \ ^3S$  &   $ 6.7391 $   &            \\[-0.09in]
  $1s^22s2p^63s  \ ^1S$  &   $ 6.8192 $   &            \\[-0.09in]
  $1s^22s2p^63p  \ ^3P$  &   $ 7.1714 $   &            \\[-0.09in]
  $1s^22s2p^63p  \ ^1P$  &   $ 7.2026 $   & $  7.204 $ \\[-0.06in]
  \hline
  $1s2s^22p^63s  \ ^3S$  &   $95.9528 $   &            \\[-0.09in]
  $1s2s^22p^63s  \ ^1S$  &   $96.0030 $   &            \\[-0.09in]
  $1s2s^22p^63p  \ ^3P$  &   $96.3928 $   &            \\[-0.09in]
  $1s2s^22p^63p  \ ^1P$  &   $96.4199 $   &            \\[-0.05in]
  \\*[-0.3in] \hline
  \end{tabular}
  }
\end{table*}

\clearpage
\begin{table*}[!h]
  \label{mgIIIen}
  \caption{
  Energies (Ryd) of the \ion{Mg}{4} target states and \ion{Mg}{3} ground state.
  } \vspace*{0.07in}
  \resizebox{4in}{!}{
  \begin{tabular}{l r r}  
  \hline \hline
  State               & Present & NIST\\
  \hline
  $1s^22s^22p^6 \ ^1S$  & $-6.0727$  & $ -5.8955 $  \\[-0.07in]
  \hline
  $1s^22s^22p^5 \ ^2P$  & $ 0.0000$  & $  0.0000 $  \\[-0.09in]
  $1s^22s2p^6   \ ^2S$  & $ 2.6694$  & $  2.8321 $  \\[-0.06in]
  \hline
  $1s2s^22p^6   \ ^2S$  & $91.9430$                 \\[-0.05in]
  \hline
  \end{tabular}
  }
\end{table*}

\clearpage
\begin{table*}[!hbtp]
  \caption{\label{mgIVen}
  Energies (Ryd) of the \ion{Mg}{5} target states and \ion{Mg}{4} ground state.
  \vspace*{0.07in}}
  \resizebox{4in}{!}{
  \begin{tabular}{l r r}  
  \hline \hline
  State               & Present & NIST\\
  \hline
  $1s^22s^2p^5 \ ^2P$  &  $-8.2847 $  & $-8.0471 $ \\[-0.07in]
  \hline
  $1s^22s^2p^4 \ ^3P$  &  $ 0.0000 $  & $ 0.0000 $ \\[-0.09in]
  $1s^22s^2p^4 \ ^1D$  &  $ 0.3369 $  & $ 0.3194 $ \\[-0.09in]
  $1s^22s^2p^4 \ ^1S$  &  $ 0.7227 $  & $ 0.6963 $ \\[-0.09in]
  $1s^22s^2p^5 \ ^3P$  &  $ 2.3940 $  & $ 2.5803 $ \\[-0.09in]
  $1s^22s^2p^5 \ ^1P$  &  $ 3.4698 $  & $ 3.6142 $ \\[-0.09in]
  $1s^22p^6    \ ^1S$  &  $ 5.9444 $  & $ 6.0335 $ \\[-0.05in]
  \hline
  $1s2s^22p^5  \ ^3P$  &  $92.9165 $  &            \\[-0.09in]
  $1s2s^22p^5  \ ^1P$  &  $93.3603 $  &            \\[-0.09in]
  $1s2s2p^6    \ ^3S$  &  $95.3646 $  &            \\[-0.09in]
  $1s2s2p^6    \ ^1S$  &  $96.0868 $  &            \\[-0.05in]
  \hline
  \end{tabular}
  }
\end{table*}

\clearpage
\begin{table*}[!hbtp]
  \caption{\label{mgVen}
  Energies (Ryd) of the \ion{Mg}{6} target states and \ion{Mg}{5} ground state.
  \vspace*{0.07in}}
  \resizebox{4in}{!}{
  \begin{tabular}{l r r}  
  \hline \hline
  State               & Present & NIST\\
  \hline
  $1s^22s^2p^4   \ ^3P$ &  $ -10.3562 $  & $ -10.3880 $ \\[-0.07in]
  \hline
  $1s^22s^22p^3  \ ^4S$ &  $  0.00000 $  & $   0.0000 $ \\[-0.09in]
  $1s^22s^22p^3  \ ^2D$ &  $  0.52952 $  & $   0.5045 $ \\[-0.09in]
  $1s^22s^22p^3  \ ^2P$ &  $  0.79048 $  & $   0.7654 $ \\[-0.09in]
  $1s^22s2p^4    \ ^4P$ &  $  2.24151 $  & $   2.2682 $ \\[-0.09in]
  $1s^22s2p^4    \ ^2D$ &  $  3.12096 $  & $   3.1144 $ \\[-0.09in]
  $1s^22s2p^4    \ ^2S$ &  $  3.67859 $  & $   3.6617 $ \\[-0.09in]
  $1s^22s2p^4    \ ^2P$ &  $  3.90098 $  & $   3.8805 $ \\[-0.09in]
  $1s^22p^5      \ ^2P$ &  $  5.92479 $  & $   5.9482 $ \\[-0.07in]
  \hline
  $1s2s^22p^4    \ ^4P$ &  $ 93.60545 $  &              \\[-0.09in]
  $1s2s^22p^4    \ ^2D$ &  $ 94.23068 $  &              \\[-0.09in]
  $1s2s^22p^4    \ ^2P$ &  $ 94.36199 $  &              \\[-0.09in]
  $1s2s^22p^4    \ ^2S$ &  $ 94.61941 $  &              \\[-0.09in]
  $1s2s2p^5(^3P) \ ^4P$ &  $ 95.82506 $  &              \\[-0.09in]
  $1s2s2p^5(^3P) \ ^2P$ &  $ 96.70406 $  &              \\[-0.09in]
  $1s2s2p^5(^1P) \ ^2P$ &  $ 97.27956 $  &              \\[-0.09in]
  $1s2p^6        \ ^2S$ &  $ 99.41050 $  &              \\[-0.05in]
  \hline
  \end{tabular}
  }
\end{table*}

\clearpage
\begin{table*}[!hbtp]
  \caption{\label{mgVIen}
  Energies (Ryd) of the \ion{Mg}{7} target states and \ion{Mg}{6} ground state.
  \vspace*{0.07in}}
  \resizebox{4.in}{!}{
  \begin{tabular}{l r r}  
  \hline \hline
  State               & Present & NIST\\
  \hline
  $1s^22s^22p^3  \ ^4S$ & $-13.7208 $  & $-13.7260 $ \\[-0.07in]
  \hline
  $1s^22s^22p^2  \ ^3P$ & $  0.0000 $  & $  0.0000 $ \\[-0.09in]
  $1s^22s2p^3    \ ^5S$ & $  1.0016 $  & $  1.0580 $ \\[-0.09in]
  $1s^22s2p^3    \ ^3D$ & $  2.0873 $  & $  2.1044 $ \\[-0.09in]
  $1s^22s2p^3    \ ^3P$ & $  2.4774 $  & $  2.4870 $ \\[-0.09in]
  $1s^22s2p^3    \ ^3S$ & $  3.2739 $  & $  3.2817 $ \\[-0.09in]
  $1s^22p^4      \ ^3P$ & $  4.8865 $  & $  4.9331 $ \\[-0.07in]
  \hline
  $1s2s^22p^3    \ ^5S$ & $ 93.6369 $  &             \\[-0.09in]
  $1s2s^22p^3    \ ^3D$ & $ 94.4994 $  &             \\[-0.09in]
  $1s2s^22p^3    \ ^3S$ & $ 94.7250 $  &             \\[-0.09in]
  $1s2s^22p^3    \ ^3P$ & $ 94.7628 $  &             \\[-0.09in]
  $1s2s2p^4      \ ^5P$ & $ 95.4771 $  &             \\[-0.09in]
  $1s2s2p^4(^4P) \ ^3P$ & $ 96.6606 $  &             \\[-0.09in]
  $1s2s2p^4      \ ^3D$ & $ 96.6883 $  &             \\[-0.09in]
  $1s2s2p^4      \ ^3S$ & $ 97.3056 $  &             \\[-0.09in]
  $1s2s2p^4(^2P) \ ^3P$ & $ 97.5790 $  &             \\[-0.09in]
  $1s2p^5        \ ^3P$ & $ 99.2593 $  &             \\[-0.05in]
  \hline
  \end{tabular}
  }
\end{table*}

\clearpage
\begin{table*}[!hbtp]
  \caption{\label{mgVIIen}
  Energies (Ryd) of the \ion{Mg}{8} target states and \ion{Mg}{7} ground state.
  \vspace*{0.07in}}
  \resizebox{4.in}{!}{
  \begin{tabular}{l r r}  
  \hline \hline
  State               & Present & NIST\\
  \hline
  $1s^22s^22p^2 \ ^3P$  & $-16.6018$  & $-16.5380 $ \\[-0.07in]
  \hline
  $1s^22s^22p   \ ^2P$  & $  0.0000$  & $  0.0000 $ \\[-0.09in]
  $1s^22s2p^2   \ ^4P$  & $  1.0960$  & $  1.1799 $ \\[-0.09in]
  $1s^22s2p^2   \ ^2D$  & $  2.0518$  & $  2.0967 $ \\[-0.09in]
  $1s^22s2p^2   \ ^2S$  & $  2.7119$  & $  2.6981 $ \\[-0.09in]
  $1s^22s2p^2   \ ^2P$  & $  2.8492$  & $  2.8965 $ \\[-0.09in]
  $1s^22p^3     \ ^4S$  & $  3.6653$  & $  3.7490 $ \\[-0.09in]
  $1s^22p^3     \ ^2D$  & $  4.1728$  & $  4.2244 $ \\[-0.09in]
  $1s^22p^3     \ ^2P$  & $  4.7639$  & $  4.7621 $ \\[-0.07in]
  \hline
  $1s2s^22p^2   \ ^4P$  & $ 94.9623$  &             \\[-0.09in]
  $1s2s^22p^2   \ ^2D$  & $ 95.6660$  &             \\[-0.09in]
  $1s2s^22p^2   \ ^2P$  & $ 95.8214$  &             \\[-0.09in]
  $1s2s^22p^2   \ ^2S$  & $ 96.0915$  &             \\[-0.09in]
  $1s2s2p^3     \ ^4D$  & $ 96.5601$  &             \\[-0.09in]
  $1s2s2p^3     \ ^4S$  & $ 96.6315$  &             \\[-0.09in]
  $1s2s2p^3     \ ^4P$  & $ 97.0167$  &             \\[-0.09in]
  $1s2s2p^3     \ ^2D$  & $ 97.5622$  &             \\[-0.09in]
  $1s2s2p^3     \ ^4S$  & $ 97.8941$  &             \\[-0.09in]
  $1s2s2p^3     \ ^2P$  & $ 98.0187$  &             \\[-0.09in]
  $1s2s2p^3     \ ^2D$  & $ 98.1266$  &             \\[-0.09in]
  $1s2s2p^3     \ ^2P$  & $ 98.5930$  &             \\[-0.09in]
  $1s2s2p^3     \ ^2S$  & $ 98.6195$  &             \\[-0.09in]
  $1s2p^4       \ ^4P$  & $ 99.0702$  &             \\[-0.09in]
  $1s2p^4       \ ^2D$  & $ 99.7024$  &             \\[-0.09in]
  $1s2p^4       \ ^2P$  & $ 99.8864$  &             \\[-0.09in]
  $1s2p^4       \ ^2S$  & $100.6505$  &             \\[-0.05in]
  \hline
  \end{tabular}
  }
\end{table*}

\clearpage
\begin{table*}[!hbtp]
  \caption{\label{mgVIIIen}
  Energies (Ryd) of the \ion{Mg}{9} target states and \ion{Mg}{8} ground state. } \vspace*{0.07in}
  \resizebox{4in}{!}{
  \begin{tabular}{l r r}  
  \hline \hline
  State                 & Present & NIST \\
  \hline
  $1s^22s^22p\ ^2P$   & $-19.5175$ & $-19.5450$ \\[-0.07in]
  \hline
  $1s^22s^2\ ^1S$     & $ 0.0000$  & $ 0.0000 $ \\[-0.09in]
  $1s^22s2p\ ^3P$     & $ 1.2720$  & $ 1.3020 $ \\[-0.09in]
  $1s^22s2p\ ^1P$     & $ 2.4716$  & $ 2.4758 $ \\[-0.09in]
  $1s^22p^2\ ^3P$     & $ 3.3112$  & $ 3.3555 $ \\[-0.09in]
  $1s^22p^2\ ^1D$     & $ 3.6698$  & $ 3.6915 $ \\[-0.09in]
  $1s^22p^2\ ^1S$     & $ 4.6220$  & $ 4.5530 $ \\[-0.07in]
  \hline
  $1s2s^22p\ ^3P$     & $ 96.3470$ &            \\[-0.09in]
  $1s2s^22p\ ^1P$     & $ 96.9743$ & $97.1192 $ \\[-0.09in]
  $1s2s(^3S)2p^2\ ^3D$& $ 97.9274$ &            \\[-0.09in]
  $1s2s(^1S)2p^2\ ^3P$& $ 97.9575$ &            \\[-0.09in]
  $1s2s(^3S)2p^2\ ^3S$& $ 98.6507$ &            \\[-0.09in]
  $1s2s(^3S)2p^2\ ^1D$& $ 98.7677$ &            \\[-0.09in]
  $1s2s(^3S)2p^2\ ^3P$& $ 98.8326$ &            \\[-0.09in]
  $1s2s(^3S)2p^2\ ^1P$& $ 99.3433$ &            \\[-0.09in]
  $1s2s(^1S)2p^2\ ^1S$& $ 99.4907$ &            \\[-0.09in]
  $1s2p^3\ ^3D$       & $ 99.6064$ &            \\[-0.09in]
  $1s2p^3\ ^3S$       & $ 99.9059$ &            \\[-0.09in]
  $1s2p^3\ ^1D$       & $100.1887$ &            \\[-0.09in]
  $1s2p^3\ ^3P$       & $100.2962$ &            \\[-0.09in]
  $1s2p^3\ ^1P$       & $100.8812$ &            \\*[-0.05in]
  \hline
  \end{tabular}
  }
\end{table*}

\clearpage
\begin{table*}[!h]
  \label{mgIXen}
  \caption{Energies (Ryd) of the \ion{Mg}{10} target states and \ion{Mg}{9} ground state.
  } \vspace*{0.07in}
  \resizebox{4.in}{!}{
  \begin{tabular}{l r r}  
  \hline \hline
  State               & Present & NIST\\
  \hline
  $1s^22s^2\ ^1$     & $-24.0872$ & $-24.1060 $   \\[-0.06in]
  \hline
  $1s^22s\ ^2S$       & $  0.0000 $ & $ 0.0000 $   \\[-0.09in]
  $1s^22p\ ^2P$       & $  1.4855 $ & $ 1.4823 $   \\[-0.06in]
  \hline
  $1s2s^2\ ^2S$       & $ 96.4886 $ & $        $   \\[-0.09in]
  $1s2s(^1S)2p \ ^2P$ & $ 97.9741 $ & $98.1544 $   \\[-0.09in]
  $1s2s(^3S)2p \ ^2P$ & $ 98.5011 $ & $98.6810 $   \\[-0.09in]
  $1s2p^2\ ^2D$       & $ 99.0911 $ & $99.2730 $   \\[-0.09in]
  $1s2p^2\ ^2P$       & $ 99.3072 $ & $99.5100 $   \\[-0.09in]
  $1s2p^2\ ^2S$       & $100.2157 $ & $100.310 $   \\[-0.05in]
  \hline
  \end{tabular}
  }
\end{table*}

\clearpage
\begin{table*}[!hbtp]
  \caption{\label{mgXen}
  Energies (Ryd) of the \ion{Mg}{11} target states and \ion{Mg}{10} ground state.
  \vspace*{0.07in}}
  \resizebox{4.in}{!}{
  \begin{tabular}{l r r}  
  \hline \hline
  State               & Present & NIST\\
  \hline
  $1s^22s\ ^2S       $  & $-26.9543$ & $-27.0100$  \\[-0.07in]
  \hline
  $1s^2\ ^1S         $  & $  0.0000$ & $  0.0000$  \\[-0.09in]
  $1s2s\ ^3S         $  & $ 97.6959$ & $ 97.8349$  \\[-0.09in]
  $1s2p\ ^3P         $  & $ 98.5492$ & $ 98.7331$  \\[-0.09in]
  $1s2s\ ^1S         $  & $ 98.6309$ & $ 98.7702$  \\[-0.09in]
  $1s2p\ ^1P         $  & $ 99.2148$ & $ 99.3884$  \\*[0.00in]
  \hline
  \end{tabular}
  }
\end{table*}
\clearpage
\begin{table*}[!h]
\caption{\label{mgIaw}
  Present Auger widths (in Ryd) for the three \ion{Mg}{2} autoionizing
  target states above the K-shell threshold (see Table~\ref{mgIen}).
  Also shown are the level-averaged HFR1 results~\citep{palmeri08}.}
  \vspace*{0.07in} \centering
  \resizebox{\textwidth}{!}{
  \begin{tabular}{r l c c c c}  
  \hline \hline
   & State  &        & Present & & HFR1 \\
  \cline{2-2}\cline{4-4}\cline{6-6}
  1 & $1s^22s22p^63s^2          \ ^2S$  & & 2.54$\times 10^{-2}$ & & 2.39$\times 10^{-2}$ \\[-0.09in]
  2 & $1s^22s22p^63s(^1S)3p     \ ^2P$  & & 1.73$\times 10^{-2}$ & & 2.37$\times 10^{-2}$ \\[-0.09in]
  3 & $1s2s^22p^63s(^3S)3p      \ ^2P$  & & 2.26$\times 10^{-2}$ & & 2.36$\times 10^{-2}$ \\[-0.05in]
  \hline
  \\[-.4in]
  \end{tabular}
  }
\end{table*}

\clearpage
\begin{table*}[!h]
  \caption{\label{mgIIaw}
  Present Auger widths (in Ryd) for the four \ion{Mg}{3} autoionizing
  target states above the K-shell threshold (see Table~\ref{mgIIen}).
  Also shown are the level-averaged HFR1 results~\citep{palmeri08}.}
  \vspace*{0.07in} \centering
  \resizebox{\textwidth}{!}{
  \begin{tabular}{r l c c c c}  
  \hline \hline
   & State  &        & Present & & MCBP \\
  \cline{2-2}\cline{4-4}\cline{6-6}
  1 & $1s2s^22p^63s  \ ^3S$  & & 2.39$\times 10^{-2}$ & & 2.68$\times 10^{-2}$ \\[-0.09in]
  2 & $1s2s^22p^63s  \ ^1S$  & & 2.41$\times 10^{-2}$ & & 2.52$\times 10^{-2}$ \\[-0.09in]
  3 & $1s2s^22p^63p  \ ^3P$  & & 2.37$\times 10^{-2}$ & & 2.45$\times 10^{-2}$ \\[-0.09in]
  4 & $1s2s^22p^63p  \ ^1P$  & & 2.33$\times 10^{-2}$ & & 2.44$\times 10^{-2}$ \\[-0.05in]
  \hline
  \\[-.4in]
  \end{tabular}
  }
\end{table*}

\clearpage
\begin{table*}[!h]
  \caption{\label{mgIIIaw}
  Present Auger widths (in Ryd) for the \ion{Mg}{4} autoionizing
  target state above the K-shell threshold (see Table~\ref{mgIIIen}).
  Also shown are level-averaged HFR1 widths~\citep{palmeri08} and AUTOSTRUCTURE results \citep{km1}.
  }\vspace*{0.07in}
  \begin{minipage}[t]{\textwidth}
  \centering
  \resizebox{5.in}{!}{
  \begin{tabular}{r l c c c c c c}  
  \hline \hline
  & State  &        & Present & & HFR1 & & AUTO \\
  \cline{2-2}\cline{4-4}\cline{6-6}\cline{8-8}
   & $1s2s^22p^6   \ ^2S$       & &  2.45$\times 10^{-2}$ & & 2.45$\times 10^{-3}$ & & 3.01$\times 10^{-2}$\\[-0.05in]
  \hline
  \end{tabular}
  }
  \end{minipage}
\end{table*}

\clearpage
\begin{table*}[!h]
  \caption{\label{mgIVaw}
  Present Auger widths (in Ryd) for the four \ion{Mg}{5} autoionizing
  target states above the K-shell threshold (see Table~\ref{mgIVen}).
  Also shown are the level-averaged results~\citep{palmeri08}.}
  \vspace*{0.07in} \centering
  \resizebox{\textwidth}{!}{
  \begin{tabular}{r l c c c c}  
  \hline \hline
  & State  &        & Present & & MCBP \\
  \cline{2-2}\cline{4-4}\cline{6-6}
  1 & $1s2s^22p^5  \ ^3P$  & & 2.13$\times 10^{-2}$ & & 2.22$\times 10^{-2}$ \\[-0.09in]
  2 & $1s2s^22p^5  \ ^1P$  & & 2.02$\times 10^{-2}$ & & 2.06$\times 10^{-2}$ \\[-0.09in]
  3 & $1s2s2p^6    \ ^3S$  & & 1.90$\times 10^{-2}$ & & 2.00$\times 10^{-2}$ \\[-0.09in]
  4 & $1s2s2p^6    \ ^1S$  & & 2.84$\times 10^{-2}$ & & 3.05$\times 10^{-2}$ \\[-0.05in]
  \hline
  \\[-.4in]
  \end{tabular}
  }
\end{table*}

\clearpage
\begin{table*}[!h]
  \caption{\label{mgVaw}
  Present Auger widths (in Ryd) for the eight \ion{Mg}{6} autoionizing
  target states above the K-shell threshold (see Table~\ref{mgVen}).
  Also shown are the level-averaged HFR1 results~\citep{palmeri08}.}
  \vspace*{0.07in} \centering
  \resizebox{\textwidth}{!}{
  \begin{tabular}{r l c c c c}  
  \hline \hline
  & State  &        & Present & & HFR1 \\
  \cline{2-2}\cline{4-4}\cline{6-6}
  1 & $1s2s^22p^4    \ ^4P$  & & 1.66$\times 10^{-2}$ &  & 1.64$\times 10^{-2}$ \\[-0.09in]
  2 & $1s2s^22p^4    \ ^2D$  & & 2.00$\times 10^{-2}$ &  & 1.96$\times 10^{-2}$ \\[-0.09in]
  3 & $1s2s^22p^4    \ ^2P$  & & 1.45$\times 10^{-2}$ &  & 1.39$\times 10^{-2}$ \\[-0.09in]
  4 & $1s2s^22p^4    \ ^2S$  & & 1.87$\times 10^{-2}$ &  & 1.83$\times 10^{-2}$ \\[-0.09in]
  5 & $1s2s2p^5(^3P) \ ^4P$  & & 1.49$\times 10^{-2}$ &  & 1.46$\times 10^{-2}$ \\[-0.09in]
  6 & $1s2s2p^5(^3P) \ ^2P$  & & 1.97$\times 10^{-2}$ &  & 1.89$\times 10^{-2}$ \\[-0.09in]
  7 & $1s2s2p^5(^1P) \ ^2P$  & & 2.02$\times 10^{-2}$ &  & 2.03$\times 10^{-2}$ \\[-0.09in]
  8 & $1s2p^6        \ ^2S$  & & 1.89$\times 10^{-2}$ &  & 1.87$\times 10^{-2}$ \\[-0.05in]
  \hline
  \\[-.4in]
  \end{tabular}
  }
\end{table*}

\clearpage
\begin{table*}[!h]
  \caption{\label{mgVIaw}
  Present Auger widths (in Ryd) for the ten \ion{Mg}{7} autoionizing
  target states above the K-shell threshold (see Table~\ref{mgVIen}).
  Also shown are level-averaged HFR1~\citep{palmeri08}, level-averaged MCBP~\citep{km4}, and level-averaged MCDF~\citep{km4} widths.}
  \vspace*{0.07in} \centering
  \resizebox{\textwidth}{!}{
  \begin{tabular}{r l c c c c c c c c}  
  \hline \hline
  & State  &   & Present & & HFR1 & & MCDF & & MCBP \\
  \cline{2-2}\cline{4-4}\cline{6-6}\cline{8-8}\cline{10-10}
   1 & $1s2s^22p^3    \ ^5S$  & & 9.31$\times 10^{-3}$ & & 9.53$\times 10^{-3}$  & & 1.36$\times 10^{-2}$ & & 9.95$\times 10^{-3}$\\[-0.09in]
   2 & $1s2s^22p^3    \ ^3D$  & & 1.54$\times 10^{-2}$ & & 1.53$\times 10^{-2}$  & & 1.91$\times 10^{-2}$ & & 1.59$\times 10^{-2}$\\[-0.09in]
   3 & $1s2s^22p^3    \ ^3S$  & & 5.82$\times 10^{-3}$ & & 5.52$\times 10^{-3}$  & & 8.89$\times 10^{-3}$ & & 5.72$\times 10^{-3}$\\[-0.09in]
   4 & $1s2s^22p^3    \ ^3P$  & & 1.45$\times 10^{-2}$ & & 1.40$\times 10^{-2}$  & & 1.77$\times 10^{-2}$ & & 1.46$\times 10^{-2}$\\[-0.09in]
   5 & $1s2s2p^4      \ ^5P$  & & 8.57$\times 10^{-3}$ & & 8.56$\times 10^{-3}$  & &            & &           \\[-0.09in]
   6 & $1s2s2p^4(^4P) \ ^3P$  & & 1.18$\times 10^{-2}$ & & 1.15$\times 10^{-2}$  & &            & &           \\[-0.09in]
   7 & $1s2s2p^4      \ ^3D$  & & 1.37$\times 10^{-2}$ & & 1.36$\times 10^{-2}$  & &            & &           \\[-0.09in]
   8 & $1s2s2p^4      \ ^3S$  & & 1.10$\times 10^{-2}$ & & 1.08$\times 10^{-2}$  & &            & &           \\[-0.09in]
   9 & $1s2s2p^4(^2P) \ ^3P$  & & 1.50$\times 10^{-2}$ & & 1.51$\times 10^{-2}$  & &            & &           \\[-0.09in]
  10 & $1s2p^5        \ ^3P$  & & 1.46$\times 10^{-2}$ & & 1.48$\times 10^{-2}$  & &            & &           \\[-0.05in]
  \hline
  \\[-.4in]
  \end{tabular}
  }
\end{table*}

\clearpage
\begin{table*}[!h]
  \caption{\label{mgVIIaw}
  Present Auger widths (in Ryd) for the 17 \ion{Mg}{8} autoionizing
  target states above the K-shell threshold (see Table~\ref{mgVIIen}).
  Also shown are level-averaged HFR1~\citep{palmeri08},
  level-averaged MCDF~\citep{chenb}, and level-averaged MCBP~\citep{km3} widths.}
  \vspace*{0.07in} \centering
  \resizebox{\textwidth}{!}{
  \begin{tabular}{r l c c c c c c c c}  
  \hline \hline
  & State  &   & Present & & HFR1 & & MCDF & & MCBP \\
  \cline{2-2}\cline{4-4}\cline{6-6}\cline{8-8}\cline{10-10}
   1 & $1s2s^22p^2   \ ^4P$ & & 8.73$\times 10^{-3}$ & & 7.81$\times 10^{-3}$ & & 8.98$\times 10^{-3}$ & & 9.06$\times 10^{-3}$ \\[-0.09in]
   2 & $1s2s^22p^2   \ ^2D$ & & 1.30$\times 10^{-2}$ & & 1.23$\times 10^{-2}$ & & 1.27$\times 10^{-2}$ & & 1.30$\times 10^{-2}$ \\[-0.09in]
   3 & $1s2s^22p^2   \ ^2P$ & & 5.63$\times 10^{-3}$ & & 4.60$\times 10^{-3}$ & & 5.37$\times 10^{-3}$ & & 5.32$\times 10^{-3}$ \\[-0.09in]
   4 & $1s2s^22p^2   \ ^2S$ & & 1.15$\times 10^{-2}$ & & 1.00$\times 10^{-2}$ & & 1.12$\times 10^{-2}$ & & 1.11$\times 10^{-2}$ \\[-0.09in]
   5 & $1s2s(^3S)2p^3\ ^4D$ & & 8.35$\times 10^{-3}$ & & 8.71$\times 10^{-3}$ & & 8.17$\times 10^{-3}$ & &            \\[-0.09in]
   6 & $1s2s(^1S)2p^3\ ^4S$ & & 1.46$\times 10^{-3}$ & & 6.92$\times 10^{-4}$ & & 1.91$\times 10^{-3}$ & &            \\[-0.09in]
   7 & $1s2s(^3S)2p^3\ ^4P$ & & 6.40$\times 10^{-3}$ & & 6.63$\times 10^{-3}$ & & 6.28$\times 10^{-3}$ & &            \\[-0.09in]
   8 & $1s2s(^1S)2p^3\ ^2D$ & & 1.13$\times 10^{-2}$ & & 1.06$\times 10^{-2}$ & & 1.15$\times 10^{-2}$ & &            \\[-0.09in]
   9 & $1s2s(^3S)2p^3\ ^4S$ & & 7.63$\times 10^{-3}$ & & 6.97$\times 10^{-3}$ & & 8.51$\times 10^{-3}$ & &            \\[-0.09in]
   10& $1s2s(^1S)2p^3\ ^2P$ & & 9.45$\times 10^{-3}$ & & 8.45$\times 10^{-3}$ & & 9.59$\times 10^{-3}$ & &            \\[-0.09in]
   11& $1s2s(^3S)2p^3\ ^2D$ & & 1.33$\times 10^{-2}$ & & 1.30$\times 10^{-2}$ & & 1.39$\times 10^{-2}$ & &            \\[-0.09in]
   12& $1s2s(^3S)2p^3\ ^2P$ & & 1.14$\times 10^{-2}$ & & 1.07$\times 10^{-2}$ & & 1.16$\times 10^{-2}$ & &            \\[-0.09in]
   13& $1s2s(^3S)2p^3\ ^2S$ & & 2.86$\times 10^{-3}$ & & 3.71$\times 10^{-3}$ & & 3.41$\times 10^{-3}$ & &            \\[-0.09in]
   14& $1s2p^4       \ ^4P$ & & 8.38$\times 10^{-3}$ & & 6.48$\times 10^{-3}$ & & 8.73$\times 10^{-3}$ & &            \\[-0.09in]
   15& $1s2p^4       \ ^2D$ & & 1.33$\times 10^{-2}$ & & 1.01$\times 10^{-2}$ & & 1.37$\times 10^{-2}$ & &            \\[-0.09in]
   16& $1s2p^4       \ ^2P$ & & 8.34$\times 10^{-3}$ & & 6.47$\times 10^{-3}$ & & 8.58$\times 10^{-3}$ & &            \\[-0.09in]
   17& $1s2p^4       \ ^2S$ & & 9.18$\times 10^{-3}$ & & 6.72$\times 10^{-3}$ & & 9.48$\times 10^{-3}$ & &            \\[-0.05in]
   \hline
  \\[-.4in]
  \end{tabular}
  }
\end{table*}

\clearpage
\begin{table*}[!h]
  \begin{minipage}[t]{\textwidth}
  \caption{\label{mgVIIIaw}
  Present Auger widths (in Ryd) for the \ion{Mg}{9} autoionizing
  target states above the K-shell threshold (see Table~\ref{mgVIIIen}).
  Also shown are level-averaged HFR1~\citep{palmeri08}, level-averaged MCDF \citep{chenbe}, and level-averaged MBCP \citep{km1} widths. \vspace*{0.07in}} \centering
  \resizebox{5.in}{!}{
  \begin{tabular}{r l c c c c c c c c}  
  \hline \hline
  & State  &        & Present & & HFR1 & & MCDF  & & MCBP\\
  \cline{2-2}\cline{4-4}\cline{6-6}\cline{8-8}\cline{10-10}
   1  & $1s2s^22p     \ ^3P$ & &  7.44$\times 10^{-3}$  & &  7.35$\times 10^{-3}$  & & 7.61$\times 10^{-3}$  & &  7.89$\times 10^{-3}$ \\[-0.09in]
   2  & $1s2s^22p     \ ^1P$ & &  5.26$\times 10^{-3}$  & &  4.84$\times 10^{-3}$  & & 5.08$\times 10^{-3}$  & &  5.12$\times 10^{-3}$ \\[-0.09in]
   3  & $1s2s(^3S)2p^2\ ^3D$ & &  6.62$\times 10^{-3}$  & &  5.95$\times 10^{-3}$  & & 5.87$\times 10^{-3}$  & &             \\[-0.09in]
   4  & $1s2s(^1S)2p^2\ ^3P$ & &  8.51$\times 10^{-4}$  & &  2.30$\times 10^{-3}$  & & 2.01$\times 10^{-3}$  & &             \\[-0.09in]
   5  & $1s2s(^3S)2p^2\ ^3S$ & &  3.36$\times 10^{-3}$  & &  3.29$\times 10^{-3}$  & & 3.10$\times 10^{-3}$  & &             \\[-0.09in]
   6  & $1s2s(^3S)2p^2\ ^1D$ & &  1.12$\times 10^{-2}$  & &  1.17$\times 10^{-2}$  & & 1.23$\times 10^{-2}$  & &             \\[-0.09in]
   7  & $1s2s(^3S)2p^2\ ^3P$ & &  6.06$\times 10^{-3}$  & &  6.27$\times 10^{-3}$  & & 6.40$\times 10^{-3}$  & &             \\[-0.09in]
   8  & $1s2s(^3S)2p^2\ ^1P$ & &  2.44$\times 10^{-3}$  & &  2.16$\times 10^{-3}$  & & 1.89$\times 10^{-3}$  & &             \\[-0.09in]
   9  & $1s2s(^1S)2p^2\ ^1S$ & &  8.24$\times 10^{-3}$  & &  8.03$\times 10^{-3}$  & & 8.76$\times 10^{-3}$  & &             \\[-0.09in]
  10  & $1s2p^3       \ ^3D$ & &  8.43$\times 10^{-3}$  & &  8.80$\times 10^{-3}$  & & 8.50$\times 10^{-3}$  & &             \\[-0.09in]
  11  & $1s2p^3       \ ^3S$ & &              & &  1.14$\times 10^{-5}$  & & 1.49$\times 10^{-5}$  & &             \\[-0.09in]
  12  & $1s2p^3       \ ^1D$ & &  8.39$\times 10^{-3}$  & &  8.61$\times 10^{-3}$  & & 8.27$\times 10^{-3}$  & &             \\[-0.09in]
  13  & $1s2p^3       \ ^3P$ & &  5.32$\times 10^{-3}$  & &  5.53$\times 10^{-3}$  & & 5.31$\times 10^{-3}$  & &             \\[-0.09in]
  14  & $1s2p^3       \ ^1P$ & &  5.05$\times 10^{-3}$  & &  5.27$\times 10^{-3}$  & & 5.03$\times 10^{-3}$  & &             \\[-0.05in]
  \hline
  \end{tabular}
  }
  \end{minipage}
  \\[0.1in]
\end{table*}

\begin{table*}[!h]
\caption{\label{mgIXaw}
  Present Auger widths (in Ryd) for the \ion{Mg}{10} autoionizing
  target states above the K-shell threshold (see Table~\ref{mgIXen}).
  Also shown are level-averaged HFR1~\citep{palmeri08}, level-averaged MCDF\citep{chenhe}, and level-averaged MBCP~\citep{km2} widths.
  }\vspace*{0.07in}
  \begin{minipage}[t]{\textwidth}
  \centering
  \resizebox{5.in}{!}{
  \begin{tabular}{r l c c c c c c c c}  
  \hline \hline
   & State  & & Present & & HFR1 & & MCDF & & MCBP \\
  \cline{2-2}\cline{4-4}\cline{6-6}\cline{8-8}\cline{10-10}
  1 & $1s2s^2     \ ^2S$  & &  6.27$\times 10^{-3}$ & &  5.81$\times 10^{-3}$   & &   5.42$\times 10^{-3}$  & &  6.28$\times 10^{-3}$ \\[-0.09in]
  2 & $1s2s(^1S)2p\ ^2P$  & &  4.06$\times 10^{-3}$ & &  4.02$\times 10^{-3}$   & &   3.53$\times 10^{-3}$  & &             \\[-0.09in]
  3 & $1s2s(^3S)2p\ ^2P$  & &  2.89$\times 10^{-4}$ & &  3.21$\times 10^{-4}$   & &   5.44$\times 10^{-4}$  & &             \\[-0.09in]
  4 & $1s2p^2     \ ^2D$  & &  6.39$\times 10^{-3}$ & &  6.90$\times 10^{-3}$   & &   6.61$\times 10^{-3}$  & &             \\[-0.09in]
  5 & $1s2p^2     \ ^2P$  & &             & &  3.98$\times 10^{-5}$   & &   3.37$\times 10^{-5}$  & &             \\[-0.09in]
  6 & $1s2p^2     \ ^2S$  & &  9.41$\times 10^{-4}$ & &  1.05$\times 10^{-3}$   & &   1.10$\times 10^{-3}$  & &             \\[-0.05in]
  \hline
  \end{tabular}
  }
  \end{minipage}
\end{table*}

\clearpage
\begin{deluxetable}{ccc}
\tablecaption{{\em XMM-Newton} RGS observations used in this paper \label{obs}}
\tablewidth{0pt}
\tablehead{
\colhead{ObsID} & \colhead{Date} & \colhead{Exposure (ks)}
}
\startdata
0150390101  &  2003 April 6  &106 \\
0150390301  &  2003 April 8  &91.5 \\
\enddata
\end{deluxetable}

\clearpage
\begin{deluxetable}{lll}
\tablecaption{GS1826-238 Mg Edge Fit \label{edge_fit}}
\tablewidth{0pt}
\tablehead{
 \colhead{Parameter}&  \colhead{{\tt TBnew}} & \colhead{{\tt warmabs}}
}
\startdata
$N_{\rm H}(10^{21}$~cm$^{-2})^{a}$        &$1.68$                  &$1.68$                 \\
A$_{\rm Mg}^{b}$                          &$5.24^{+1.68}_{-1.52}$  &$4.51^{+2.11}_{-1.02}$  \\
$\log\xi$                                 &                        &$-2.38^{+0.81}_{-0.97}$ \\
$\chi^2$                                  &$0.974$                 &$0.974$                \\
\enddata
\tablenotetext{a}{Fixed to the $21$~cm value \citep{kal05}.}
\tablenotetext{b}{Abundances relative to the solar values of \citet{lod03}.}
\end{deluxetable}

\begin{deluxetable}{lcc}
\tablecaption{Mg Ionic Column Densities \label{col_den}}
\tablewidth{0pt}
\tablehead{
  \colhead{Ion} &  \colhead{Ionization Fraction} &  \colhead{Column Density}
}
\startdata
\ion{Mg}{1} & $0.24\pm 0.04 $ & $6.43  \pm  1.28$ \\
\ion{Mg}{2} & $0.26\pm 0.05 $ & $7.08  \pm  1.41$ \\
\ion{Mg}{3} & $0.48\pm 0.09 $ & $12.98 \pm  2.51$ \\
\enddata
\tablecomments{Mg column densities in units of $10^{16}$~cm$^{-2}$. Values are derived from the {\tt warmabs} fit using
the Mg solar abundance in \citet{lod03}.}
\end{deluxetable}

\end{document}